\documentclass[10pt,preprint]{aastex}
\usepackage{epsf}
\usepackage{amsmath}

\newcommand{\spitzer}{\textit{Spitzer}}

\newcommand{\rsdss}{\hbox{$r$}}
\newcommand{\isdss}{\hbox{$i$}}

\newcommand{\hb}{\hbox{H$\beta$}}
\newcommand{\ha}{\hbox{H$\alpha$}}
\newcommand{\lsim}{\lesssim}
\newcommand{\gsim}{\gtrsim}

\newcommand{\etal}{et al.}
\newcommand{\eg}{e.g.}

\newcommand{\uJy}{\hbox{$\mu$Jy}}
\newcommand{\ujy}{\hbox{$\mu$Jy}}

\newcommand{\cmodel}{\hbox{\tt cmodel}}

\newcommand{\tabletwo}{
\tablewidth{0pt}
\tablecaption{Target Hectospec Fields\label{table:obs}}
\tablehead{ \colhead{} & \colhead{R.A.} & \colhead{Decl.} &
\colhead{$t_\mathrm{exp}$} & \colhead{Observation Date} & \colhead{} \\
\colhead{Field} & \colhead{(J2000)} & \colhead{(J2000)} &
\colhead{(min)} & \colhead{(UTC)} & \colhead{Airmass}}
\startdata
131 & 17\phn14\phn19.5 & +59\phn51\phn29 &  45 & 2004 Jun 13 & 1.23\\
132 & 17\phn19\phn59.1 & +59\phn52\phn10 &  45 & 2004 Jun 13 & 1.35\\
133 & 17\phn12\phn54.9 & +59\phn03\phn24 &  45 & 2004 Jun 15 & 1.23\\
134 & 17\phn18\phn16.6 & +59\phn01\phn48 &  45 & 2004 Jun 20 & 1.16\\
135 & 17\phn23\phn57.6 & +58\phn55\phn59 &  45 & 2004 Jun 23 & 1.26\\
\enddata
\tablecomments{Units of right ascension are hours, minutes, and seconds,
and units of declination are degrees, arcminutes, and
arcseconds.}
}

\newcommand{\tableone}{
\tablewidth{0pt}
\tablecaption{Main SExtractor Parameters for the 24~\micron\
Catalog\label{table:sex}} 
\tablehead{ \colhead{Parameter} & \colhead{Value}}
\startdata
\texttt{DETECT\_MINAREA} & 3.0 \\
\texttt{DETECT\_THRESH} & 2.5 \\
\texttt{FILTER} & No \\
\texttt{DEBLEND\_NTHRESH} & 64 \\
\texttt{DEBLEND\_MINCONT} & 0.02 \\
\texttt{CLEAN} & Yes \\
\texttt{CLEAN\_PARAM} & 1.0 \\
\texttt{BACK\_SIZE} & 50 \\
\texttt{BACK\_FILTERSIZE} & 1 \\
\enddata
}

\newcommand{\figcapzhist}{Redshift distribution of 24~\micron\ sources
from the combined primary, secondary, and tertiary samples in the
\spitzer\ FLS.  In each panel, the shaded histogram shows the redshift
distribution (as labeled) of sources spectroscopically classified as
``Galaxies'', and ``QSOs'' from the Hectospec and SDSS data. The solid
line shows the redshift distribution of all the MIPS samples from the
Hectospec and SDSS samples.
\label{fig:zhist}}

\newcommand{\figcapselect}{Distribution of 24~\micron\ flux densities
of FLS sources as a function of SDSS \isdss--band magnitude.  We use
contours where more than 16 data points fall in bins of $\Delta \isdss
= 0.5$~mag and $\Delta (\log f_\nu[24\micron]) = 0.2$~dex, and we plot
individual points otherwise. The first contour corresponds to a source
density of 16 objects in these bins, and each contour corresponds to a source
density successively higher by a factor of two.   Shaded rectangles show
target selection for spectroscopy.  The red--shaded region shows
the primary selection, $f_\nu(24\micron) \geq 1$~mJy and $\isdss \leq
21$~mag.  The blue--shaded region shows the secondary selection, $0.3
\leq f_\nu(24\micron)/$1~mJy$ < 1$ and $\isdss \leq 20.5$~mag.  The
magenta--shaded region shows the tertiary selection, $f_\nu(24\micron)
< 0.3$~mJy and $\isdss \leq 20.5$~mag. Sources fainter than the $\isdss
\gsim 21.3$~mag will be missed in SDSS.\label{fig:f24vi}}

\newcommand{\figradec}{Hectospec field and SDSS--matched 24~\micron\ source
locations in the \spitzer\ FLS.  Large shaded circles show the
positions of our Hectospec fields, also listed in
table~\ref{table:obs}.  The black dots in the top panel show objects
from all three MIPS samples.  Large, boxes show 24~\micron\ sources
with successfully measured extragalactic redshifts from Hectospec
only, yellow stars show sources spectroscopically classified as
Galactic stars. The bottom panels show only those objects in the
primary (1076 objects) and secondary samples (2296 objects), as
labeled.  In the bottom panels, the large, solid boxes show the 794
primary and 693 secondary sources with redshifts from either Hectospec
or SDSS. \label{fig:radec}}

\newcommand{\figcapcomp}{The completeness of the SDSS--matched 24~\micron\
spectroscopic sample as a function of 24~\micron\ flux density.  The
filled histograms show the completeness fraction of SDSS--matched
24~\micron\ sources with successfully measured redshifts (either from
Hectospec or SDSS) to the total number of sources in each sample in
the Hectospec region.  The unfilled histogram shows the
redshift--success rate, defined as the number of sources with
successfully measured redshifts to the number of sources observed by
Hectospec in each sample.  The vertical dashed line delineates the
primary and secondary samples (as labeled).  The tertiary sample has
low completeness ($<$3\%), and we do not show it here. \label{fig:comp}}

\newcommand{\figcapnumber}{Number distribution of 24~\micron\
sources.  The filled histogram shows the SDSS--matched 24~\micron\
sample in flux--density bins of 0.1~dex.  The unfilled histogram shows
the number distribution of all 24~\micron\ sources in the FLS that
fall in the Hectospec fields.  For comparison, the solid squares show
the expected number distribution for 24~\micron\ sources based on the
total number counts from \citet{pap04}, normalized to the area of the
Hectospec survey (3.3~deg$^2$).  The drop in the number of 
SDSS--matched spectroscopic targets below $\sim$1~mJy occurs because
of the optical SDSS magnitude limit (see figure~\ref{fig:f24vi}).  The
SDSS--matched sample excludes 40\% of the 24~\micron\ sources with
$f_\nu(24\micron) > 1$~mJy.  Of these, $\sim$40\% are saturated
stars and objects too close to saturated stars, nearby galaxies
split into multiple 24~\micron\ sources, and objects with a large offset
exists between the optical and 24~\micron\ source.  The remaining
objects (25\% of the $f_\nu(24\micron) \geq 1$~mJy population) have
fainter optical counterparts, $\isdss \gsim 21$~mag. \label{fig:number}}

\newcommand{\figcapsamplespec}{Example spectra from the Hectospec
data, illustrating the problem with the atmospheric dispersion
corrector.  Each panel shows the flux--calibrated spectrum (heavy,
black line; resampled to $R \sim 400$ and interpolated over pixels
with bad mask values)  for a source denoted by its Hectospec field and
fiber number.  The red, shaded region denotes the 1$\sigma$
uncertainty.   The heavy vertical lines denotes the observed
wavelength of the [\ion{O}{2}] $\lambda$3727 line.  The panels are
ordered from left to right with increasing airmass at the time the
field was observed (see table~\ref{table:obs}).  The increase in flux
density at the blue end of the spectrum illustrates the ADC--induced
problem with the flux calibration.\label{fig:adc}}

\newcommand{\figcapmaxz}{Objects classified as a galaxy and QSO with
the highest redshift in the Hectospec data.  In each panel, the black,
solid line shows the reduced, flux--calibrated Hectospec spectrum
(resampled to $R\sim 350$), and the red, shaded region shows the
1$\sigma$ uncertainty.  Locations of prominent features are
labeled. Redshift, 24~\micron\ flux density, and \isdss--band
magnitude are inset in each panel.\label{fig:maxz}}

\shorttitle{A MMT/HECTOSPEC REDSHIFT SURVEY OF 24~MICRON SOURCES
IN THE \textit{SPITZER} FLS}
\shortauthors{PAPOVICH ET AL.}

\begin{document}

%\slugcomment{\it Revised Draft Version \today, \ampmtime}
\slugcomment{\it Accepted for Publication in the Astronomical Journal}
\title{A MMT/HECTOSPEC REDSHIFT SURVEY OF 24~MICRON SOURCES
IN THE \textit{SPITZER} FIRST LOOK SURVEY\altaffilmark{1}} 

%TRUEMODE\newif\iffullauthornames
%TRUEMODE\fullauthornamestrue
%fullauthornamesfalse

%TRUEMODE\iffullauthornames
\author{\sc Casey~Papovich\altaffilmark{2},
Richard~Cool, 
Daniel~Eisenstein,
Emeric~Le~Floc'h\altaffilmark{3},
Xiaohui~Fan,
Robert~C.~Kennicutt, Jr.\altaffilmark{4},
J.~D.~T.~Smith, 
G.~H.~Rieke
and Marianne~Vestergaard
}
%TRUEMODE\else
%TRUEMODE\author{\sc C.~Papovich\altaffilmark{2}, 
%TRUEMODER.~Cool,
%TRUEMODED.~Eisenstein,
%TRUEMODEE.~Le~Floc'h\altaffilmark{3},
%TRUEMODEX.~Fan,
%TRUEMODER.~C.~Kennicutt, Jr.\altaffilmark{4},
%TRUEMODEJ. D.~T.~Smith, and G.~H.~Rieke
%TRUEMODEand M.~Vestergaard
%TRUEMODE} 
%TRUEMODE\fi

\affil{Steward Observatory, University of Arizona, 933 North
Cherry Avenue, Tucson, AZ 85721}

\altaffiltext{1}{This work is based in part on observations
made with the \textit{Spitzer Space Telescope}, which is operated by
the Jet Propulsion laboratory, California Institute of Technology,
under NASA contract 1407}
\altaffiltext{2}{Spitzer Fellow; papovich@as.arizona.edu}
\altaffiltext{3}{Also associated to Observatoire de Paris, GEPI, 92195
Meudon, France}
\altaffiltext{4}{Current address: Institute of Astronomy, University
of Cambridge, Madingley Road, Cambridge CB3 OHA, United Kingdom}

%%%%%%%%%%%%%%%%%%%%%%%%%%%%%%%%%%%%%%%%%%%%%%%%%%%%%%%%%%%%%%%%%%%%%%

\setcounter{footnote}{4}

\begin{abstract}
We present a spectroscopic survey using the MMT/Hectospec fiber
spectrograph of 24~\micron\ sources selected with the \spitzer\ Space
Telescope in the \spitzer\ First Look Survey.  We report 1296 new
redshifts for 24~\micron\ sources, including 599 with
$f_\nu(24\micron) \geq 1$~mJy.   Combined with 291 additional
redshifts for sources from the Sloan Digital Sky Survey (SDSS), our
observing program was highly efficient and is $\sim$90\% complete for
$\isdss \leq 21$~mag and $f_\nu(24\micron) \geq 1$~mJy, and is 35\%
complete for $\isdss \leq 20.5$~mag and 0.3~mJy $\leq f_\nu(24\micron)
<$1.0~mJy.  Our Hectospec survey includes 1078 and 168 objects
spectroscopically classified as galaxies and QSOs, respectively.
Combining the Hectospec and SDSS samples, we find 24~\micron--selected
galaxies to $z_\mathrm{gal}\leq 0.98$ and QSOs to $z_\mathrm{QSO}\leq
3.6$, with mean redshifts of $\langle z_\mathrm{gal} \rangle$=0.27 and
$\langle z_\mathrm{QSO} \rangle$=1.1.  
As part of this publication, we
include the redshift catalogs and the reduced spectra; these are
also available through the NASA/IPAC Infrared Science
Archive.\footnote{http://irsa.ipac.caltech.edu/\label{footnote:url}}
\end{abstract}
 
\keywords{
galaxies: high-redshift ---
infrared: galaxies
}
 
%%%%%%%%%%%%%%%%%%%%%%%%%%%%%%%%%%%%%%%%%%%%%%%%%%%%%%%%%%%%%%%%%%%%%%

\section{Introduction}\label{section:intro}

Observations with the Infrared (IR) Astronomical Satellite
(\textit{IRAS}) discovered that much of the bolometric emission
associated with star--formation and active--galactic nuclei occurs in
the thermal infrared.  The analysis of \textit{IRAS} sources indicated
that most ($\simeq$70\%) of the light emitted from local, normal
galaxies comes at UV and optical wavelengths \citep[\eg][]{soi91}.
However, measurements of the IR background found that the total
far--IR emission ($\lambda = 8-1000$~\micron) of galaxies is
comparable to that measured at UV and optical wavelengths
\citep[\eg,][]{hau98}.  Therefore, over the history of the Universe,
roughly half of the photons from star formation or black--hole
accretion processes are emitted at IR wavelengths
\citep{elb02,dol06}.  Subsequent studies of IR number counts from the
Infrared Space Observatory (\textit{ISO}; Elbaz et al.\ 1999) and
more recently from the \spitzer\ Space Telescope \citep{mar04,pap04}
showed that IR--luminous galaxies have evolved rapidly, implying that
they are a much more common phenomenon at high redshifts.

Studying the increase in the IR--active phases of galaxies requires
measuring the properties of these objects as a function of redshift.
Observations at 24~\micron\ from the multiband imaging photometer for
\spitzer\ \citep[MIPS,][]{rie04} are particularly well suited for such
studies.  \citet{soi87} concluded that starburst
galaxies radiate as much as $\sim$40\% of their luminosity in the
mid--IR (8--40~\micron). The mid--IR emission from starforming
galaxies correlates almost linearly with total IR luminosity over a
range of galaxy type
\citep[\eg,][]{spi95,rou01,pap02,cal05}.   The angular resolution of
\spitzer\ at 24~\micron\ is roughly a factor of 3 and 7 better than
that at 70 and 160~\micron, respectively, allowing unambiguous source
identification and probing the IR emission from many more sources than
at the longer wavelengths.   Already, early studies of \spitzer\
24~\micron\ sources with photometric redshifts over relatively small
fields ($\lsim 0.5$~sq.\ deg) indicate that the bright end of the IR
luminosity function evolves strongly from $z\sim0$ to 1
\citep{lef05,per05}.

To improve our understanding of the nature and evolution of
IR-luminous phases of galaxies, we first need to construct large
samples of objects with spectroscopic redshifts. In this paper, we
publish the results of our survey with the MMT/Hectospec multi-fiber
spectrograph in the \spitzer\ First Look Survey (FLS).  We report new
redshifts for 1296 objects selected in the \isdss--band and at
24~\micron\ over 3.3~deg$^2$.   Here we publish the catalogs and
reduced, flux--calibrated spectra.   We are currently using these
spectroscopic data in conjunction with surveys in other fields to
study the evolution of the IR--luminous galaxy population.

We organize this paper as follows. In \S~\ref{section:data}, we
discuss the \spitzer\ FLS dataset, our 24~\micron\ catalog, and our
spectroscopic target selection.  In \S~3 we describe the spectroscopic
observations and data reduction.   In \S~4 we discuss the
spectroscopic completeness of the catalog.  In \S~5  we present the
Hectospec spectra and redshift catalog.  In \S~6 we summarize our
results and discuss the redshift distribution of  our sample of
24~\micron\ sources.  All magnitudes in this paper correspond to the
AB system \citep{oke83}, where $m_\mathrm{AB} = 23.9 - 2.5 \log (
f_\nu / 1\;\mu\mathrm{Jy} )$.

\section{\spitzer\ First Look Survey}\label{section:data}

\subsection{Overview} 

The \spitzer\ FLS was a service to the \spitzer\ user community,
initiated as a Director's Discretionary Time program.  The program
goal was to provide data over large areas in time to have an impact on
early \textit{Spitzer} studies and proposals.  The FLS includes three
components, described at http://ssc.spitzer.caltech.edu/fls.  Here, we
focus on the extragalactic component, whose field was chosen to have
low Galactic background and to be in the \spitzer\ continuous viewing
zone (CVZ) such that it would be observable shortly after the
\spitzer\ in-orbit checkout regardless of launch date.

The \spitzer\ FLS overlaps with imaging and spectroscopic observations
from the Sloan Digital Sky Survey \citep[SDSS;][]{yor00}.  We use the
SDSS imaging data to identify objects in the \spitzer/MIPS data for
spectroscopic follow-up.

\subsection{MIPS 24~\micron\ Observations and Data}\label{section:mipscatalog}

\spitzer\ observed the FLS during December 2003  (PID: 26; PI:
B.~T.~Soifer).  The MIPS observations consisted of medium--speed scan
maps (AORKEYS: 3863808, 3864064, 3864320, 3864576, 3864832, 3865088,
3865344, 3865600, 3865856, 3866112, 3866368, 3866624). These
24~\micron\ data cover roughly a field of 4.4~deg$^2$ at a depth of
90~sec (the ``Shallow'', or ``Main'' field).  Subsequent deeper
observations (450~sec) were taken over a smaller field (0.26~deg$^2$;
the ``Verification'' field), but those data were not available prior
to our MMT/Hectospec run.   Thus, we use only the shallower--depth data
over the larger field of view.   See \citet{mar04} for more details of
the MIPS 24~\micron\ dataset.

We retrieved the raw \spitzer/MIPS 24~\micron\ data from the \spitzer\ archive
and reduced them using the Data Analysis Tool (DAT) designed by the
MIPS Guaranteed Time Observers (GTOs; Gordon et al.\ 2005).  The
measured count rates were corrected for dark current, cosmic rays, and
flux nonlinearities, and then normalized by flat fields appropriate
for each MIPS scan--mirror position.  Images were then mosaicked after
correcting for geometric distortion with a final plate scale of
$1\farcs247$~pixel$^{-1}$.

%\subsection{MIPS 24~\micron\ Catalog}\label{section:catalog}

We constructed a catalog from the reduced and mosaicked 24~\micron\
image using SExtractor \citep{ber96}; table~\ref{table:sex} gives the
main parameters.  We measured photometry in circular apertures of
radius $12\farcs3$.  To convert these to total count rates, we applied
an aperture correction of 1.172 based on a curve--of--growth analysis
of the 24~\micron\ point spread function (PSF), and we converted these
to flux density using the current calibration factor.\footnote{See
http://ssc.spitzer.caltech.edu/mips/ for aperture corrections and
calibration factors.}  For our primary and secondary spectroscopic
targets, we selected the relatively bright MIPS sources with
$f_\nu(24\micron) \geq 300$~\uJy\ (see \S~\ref{section:select}) and
required that sources come from regions of the image with exposure
time greater than 80~sec.   Aperture--photometry for these bright
24~\micron\ objects is fairly robust.   Using aperture photometry from
SExtractor, \citet{mar04} demonstrated that the scatter between
24~\micron\ flux densities for objects in the overlap region between
the FLS shallow and verification fields is less than $\approx$20\%.
Our tests on the shallow FLS 24~\micron\ image show that in apertures
of radius $12\farcs3$ the 1$\sigma$ flux uncertainty is
$\simeq$120~\ujy.  Therefore, the photometric measurements for objects
with $f_\nu(24\micron) > 1$~mJy have high signal--to--noise ratios
(S/N$\gsim 8$), although photometric uncertainties on objects with
$f_\nu(24\micron) \sim 0.3$~mJy are higher (S/N$\sim$2--3).    Note
that the signal--to--noise ratios are significantly higher in the
(smaller) isophotal detection apertures.   The flux uncertainty of
$\sim$120~\ujy\ discussed above is only on the measurement in the
larger photometric apertures.

The depth of the MIPS 24~\micron\ data for the FLS is comparable to
those of the NOAO deep wide--field survey (NDWFS) in the constellation
Bo\"otes \citep[\eg,][]{hou05}, for which the 24~\micron\
data is roughly 80\% complete for sources above 270~\uJy\
\citep{pap04}. Similarly, \citet{mar04} estimate the 80\% completeness
for the FLS 24~\micron\ data at 230~\uJy.  We expect a
similar completeness level in the FLS 24~\micron\ catalog used here.
However, both Marleau et al.\ and Papovich et al.\ used aperture
photometry weighted by the 24~\micron\ PSF, providing more accurate
photometry vis--\'a--vis simple aperture photometry especially for
fainter sources  (particularly, e.g., at $f_\nu(24\micron) \lsim
300$~\ujy; see Marleau et al.).  Therefore, we expect more scatter in
the catalog used here particularly at faint flux densities, reducing
the overall catalog completeness (see also
\S~\ref{section:select}).   For this reason and owing to the higher
photometric uncertainties discussed above, the accuracy of the
24~\micron\ flux densities below $\lsim 0.5$~mJy may be insufficient
for some applications.

\subsection{Spectroscopic Target Selection}\label{section:select}

We selected MIPS 24~\micron\ sources from our catalog as spectroscopic
targets using the SDSS \isdss-band photometry \citep{sto02}.   We
matched the MIPS 24~\micron\ sources to SDSS \isdss--band objects from Data
Release 2 \citep[rerun 40--44,][]{aba04} within a radius of 2\arcsec\
down to $\isdss \leq 21$~mag,\footnote{The SDSS is 95\% complete at
$\isdss = 21.3$~mag \citep{aba04}.  However, we matched 24~\micron\
sources against SDSS sources selected in \textit{any} band. In
particular, the \rsdss\ band data is 95\% complete to
\rsdss$\leq$22.2~mag \citep{aba04}, allowing robust faint object
detection somewhat below the fiducial \isdss--band 95\% completeness
limit, see \eg, figure~\ref{fig:f24vi}.} excluding objects saturated
in SDSS.   We used the combined--model (\cmodel) magnitude for our $i$-band
photometry.  The SDSS fits seeing--convolved elliptical de Vaucouleurs
and exponential models to the deblended images.  The
\cmodel\ aperture goes one step further by finding the best
non-negative linear combination  of the two (de Vaucouleurs and
exponential) best fit models.  The parameter of this mixture is listed
as {\tt frac\_deV} in the SDSS data releases.  The \cmodel\ flux is
taken as this linear combination of the best--fit model fluxes.
Tests have shown that this combination tracks the Petrosian magnitudes
well but with improved performance for fainter and smaller objects
\citep{aba04}.

The SDSS deblender algorithm does not typically separate sources
closer than $\sim$2--4\arcsec\ \citep[\eg,][]{pin03}.  Although the MIPS
24~\micron\ PSF is $6\arcsec$ FWHM, source centroiding is accurate to
$<$1\arcsec.  Therefore, we expect few cases where multiple SDSS
sources lie within 2\arcsec\ of a 24~\micron\ source.  In fact, we
find that there are only 10 MIPS sources with 2 SDSS associations
($<$0.3\% of the matched 24~\micron--SDSS sample; no MIPS source has
more than 2 associations in SDSS).  In these cases, we associate the
24~\micron\ source with both SDSS sources in our catalogs for
completeness.  In practice, we placed Hectospec fibers on only 2 of
these 20 SDSS galaxies, both of which are Galactic stars.  Therefore,
the few 24~\micron\ sources with multiple object associations in SDSS
has a tiny impact on our survey.   However, there may still be some
cases where the 24~\micron--matched SDSS source is not the true
association (i.e., the true optical counterpart is fainter than the
SDSS detection limit).  Alternatively, it may also be that the
separation between the SDSS optical counterpart and the 24~\micron\
source is $>$2\arcsec, as may be the case for nearby galaxies where the optical
centroid is offset from the 24~\micron\ centroid.

To target objects for spectroscopy, we constructed a three--tiered sample:
\begin{equation}\label{eqn:sample}
\begin{array}{lrcl}
\mathrm{Primary~\;Sample,} &\;\; f_\nu(24\micron) \geq
1.0\mathrm{~mJy} & \mathrm{and} & \isdss \leq 21\mathrm{~mag}; \\ 
\mathrm{Secondary~\;Sample,} &\;\;\; 0.3\mathrm{~mJy} \leq f_\nu(24\micron) <
1.0\mathrm{~mJy} & \mathrm{and} &   \isdss \leq 20.5\mathrm{~mag}; \\ 
\mathrm{Tertiary~\;Sample,} &\;\; f_\nu(24\micron) < 0.3\mathrm{~mJy}
& \mathrm{and} &  \isdss \leq 20.5\mathrm{~mag}.
\end{array}
\end{equation}
We exclude all sources with SDSS 3\arcsec--aperture fiber magnitudes
$\rsdss \leq 17$~mag; this removes stars brighter than about
$\rsdss<17.3$ but affects very few galaxies.   This sample selection
identified 1076 sources in the primary sample, 2296 sources in the
secondary sample, and 2325 sources in the tertiary sample,
corresponding to survey densities of 0.07, 0.14, and
0.15~arcmin$^{-2}$ within the FLS shallow field, respectively.  This
is our parent sample for spectroscopy.  Table 2 presents the
astrometry and photometry from the SDSS and MIPS 24~\micron\ data for
the parent sample.   Note that each object in the parent sample is
assigned a unique ID number, in column (1) of table~\ref{table:catalog}.
These ID numbers are used in subsequent tables to identify the object
in the spectroscopic catalogs.  The target--flag values (column 9)
identifies objects satisfying the primary ({\tt Target Flag}=2),
secondary ({\tt Target Flag}=4), or tertiary sample ({\tt Target
Flag}=8), as well as calibration stars ({\tt Target Flag}=1), and
other objects  ($z$$\sim$0.5 red galaxies, and quasars; {\tt Target
Flag}=0).  Note that objects in the tertiary sample and those with
{\tt Target Flag}=0 are primarily used to fill out the fiber
configurations, and in general these samples are too sparse to be
statistically useful.

\begin{figure}
%\plotone{f24vi.ps}
\plotone{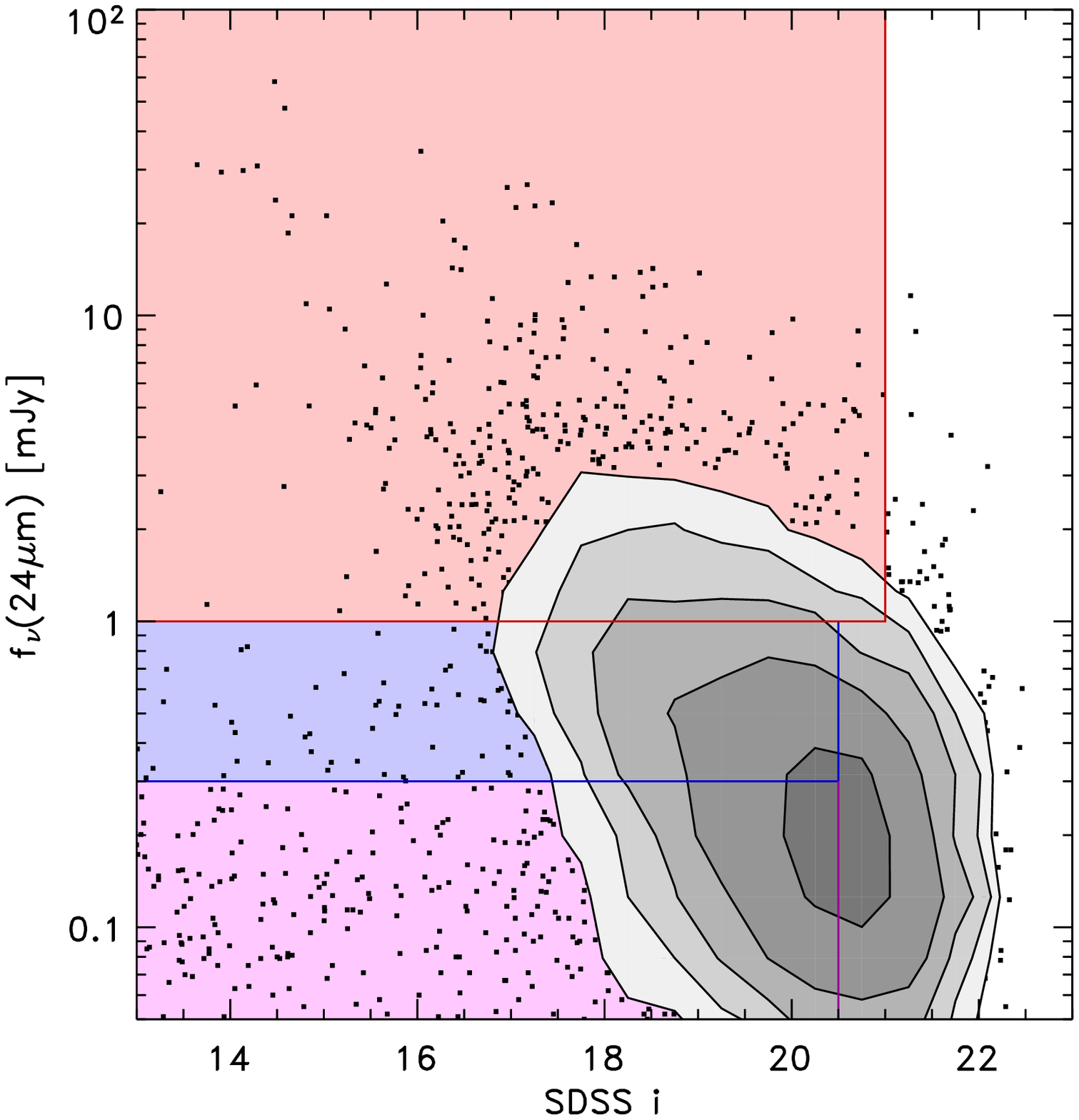}
\caption{\figcapselect}
\end{figure}

In figure~\ref{fig:f24vi} we illustrate our sample selection on a
\isdss--band magnitude versus 24~\micron\ flux--density diagram.  For
this plot, we show all MIPS 24~\micron\ sources with \isdss--band
counterparts, including sources with \isdss$>$21~mag in SDSS.  The
optical magnitude limit of our primary sample selection includes about
60\% of all 24~\micron\ sources with flux densities above 1~mJy.  We
find that saturated stars (and objects blended with the light from
saturated stars) account for 10\%  of sources with $f_\nu(24\micron) \geq
1$~mJy, and a further 5\% are galaxies where there exists a large
centroiding offset between the optical and 24~\micron\ source or where
the galaxy is split into multiple 24~\micron\ sources.  The remaining
25\% of the $f_\nu(24\micron) \geq 1$~mJy population corresponds to
objects with optical counterparts fainter than $\isdss \gsim 21$~mag,
likely similar to the IR--luminous galaxies and AGN at higher redshift
($z$$\gsim$1) studied by \citet{hou05}, \citet{yan05}, and
\citet{wee06}.

\begin{figure}
%\plotone{number.ps}
\plotone{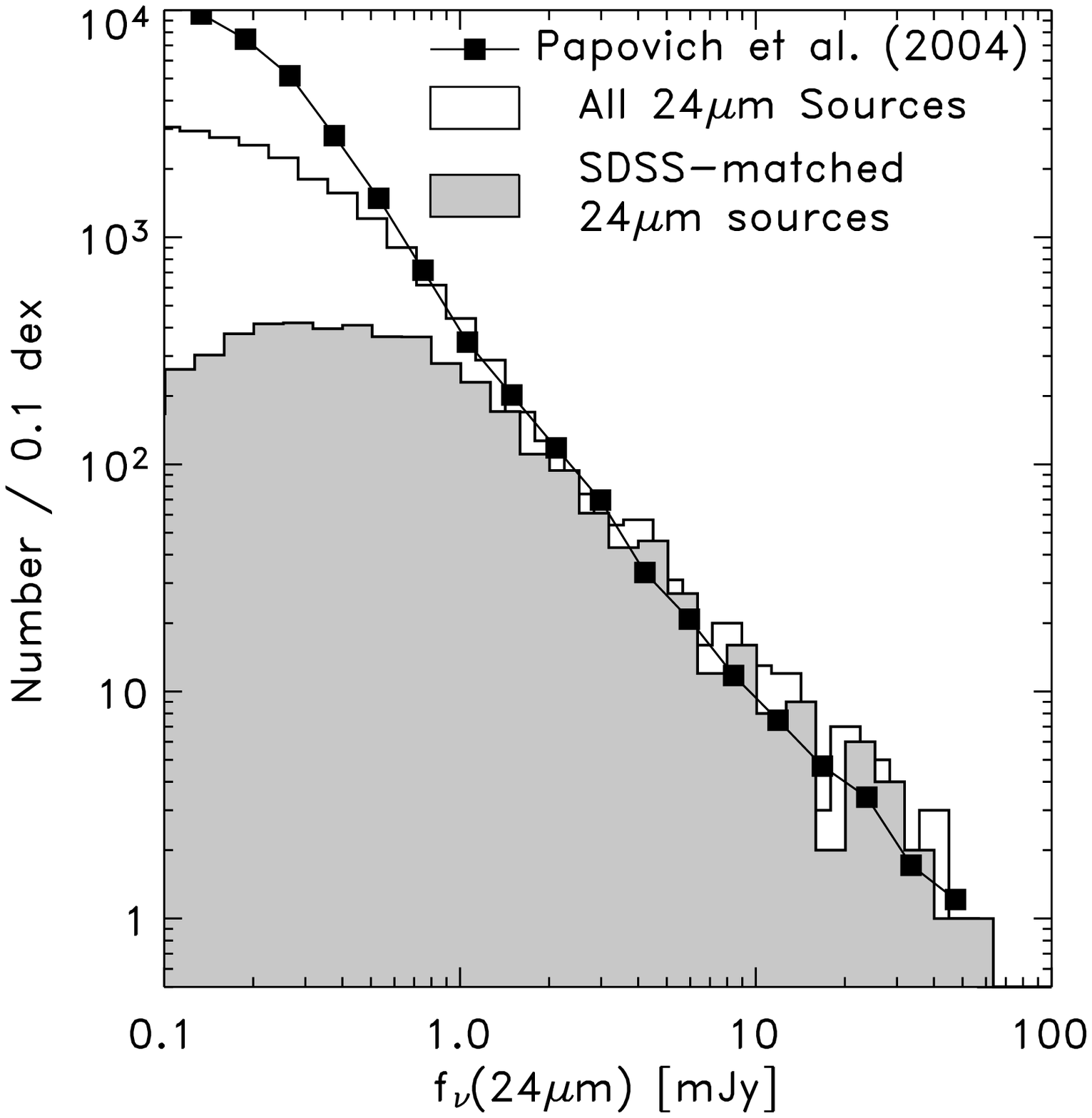}
\caption\figcapnumber
\end{figure}

The distribution of \spitzer\ 24~\micron\ sources with fainter flux
densities extends to fainter optical magnitudes.   Thus, our survey is
less complete for 24~\micron\ sources with $f_\nu(24\micron) \leq
1$~mJy --- we find that the optical magnitude limit of the secondary
sample includes $\approx$32\% of all sources with 0.3~mJy $\le
f_\nu(24\micron) <$~1~mJy in the Hectospec fields.   This
incompleteness is illustrated in figure~\ref{fig:number}, which shows
the number counts of SDSS--matched 24~\micron\ sources.  Compared with
all 24~\micron\ sources in the Hectospec fields, the number
distribution of SDSS--matched 24~\micron sources is mostly complete
for $\gsim$1~mJy, then begins to decline as the fraction of 24~\micron\
sources with SDSS--optical counterparts decreases.  As a comparison,
we show the expected number distribution of 24~\micron\ sources from
the total number counts of \citet{pap04}.

SDSS provides spectroscopic coverage over the FLS field for objects
with $\rsdss < 17.77$~mag.  From the SDSS-matched 24~\micron\
catalog, 291 objects have spectroscopic redshifts from SDSS with no
warnings on the redshift measurement ({\tt ZWARNING}=0), split between
stars (11), galaxies (223), and QSOs (57).  We list the properties of
these objects in table~\ref{table:sdsscatalog}.  ID numbers in
column~1 of table~\ref{table:sdsscatalog} correspond to the ID numbers
in column~1 of table~\ref{table:catalog}.  Because the objects in
table~\ref{table:sdsscatalog} have high--quality redshifts and spectra
from SDSS, we did not reobserve them with Hectospec.

%TRUEMODE\ifsubmode
\begin{figure}
\epsscale{0.5}
%\vbox{\plotone{radec.ps}}
\vbox{\plotone{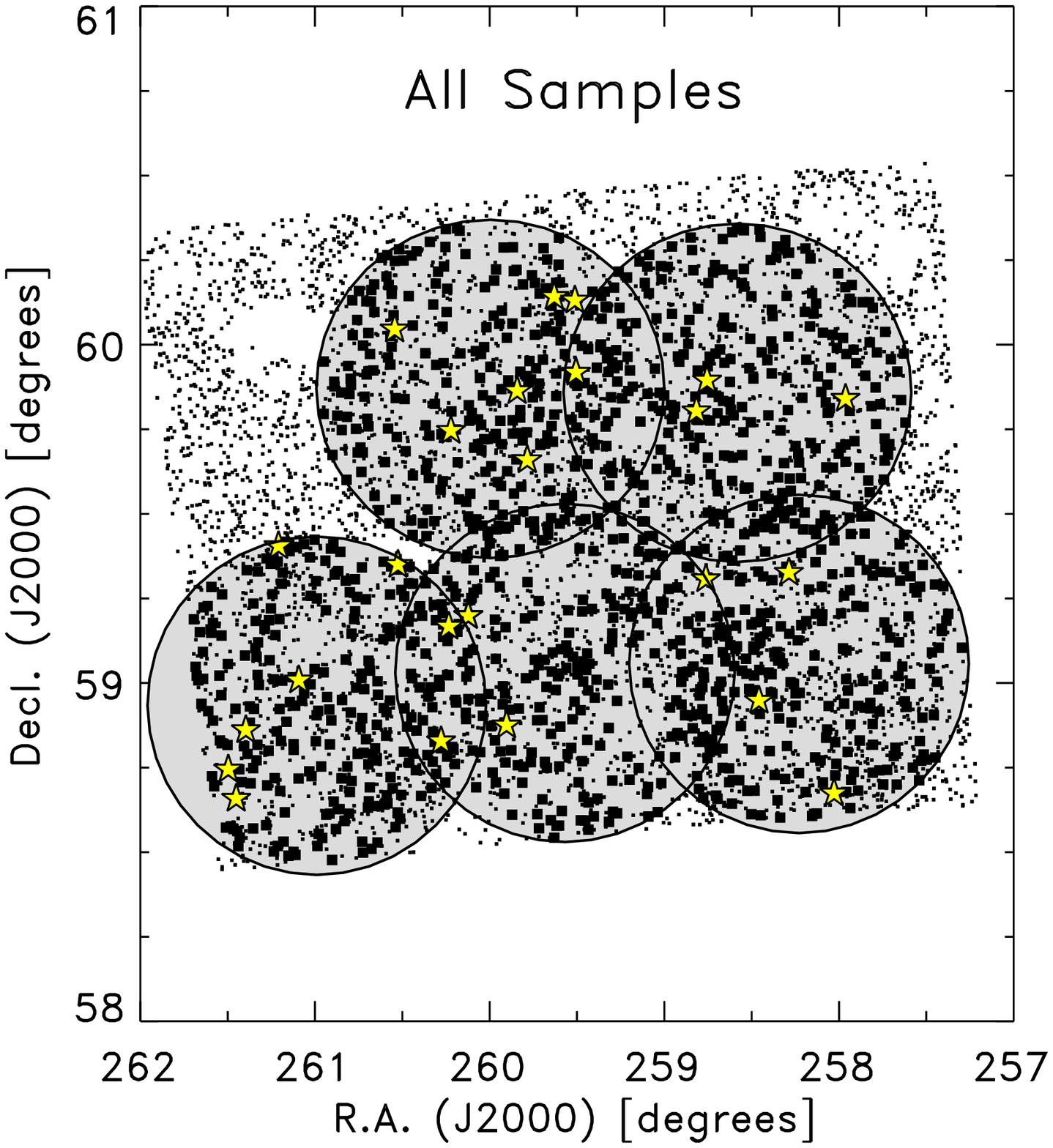}}
\epsscale{1.0}
%\vbox{\plottwo{radec_primary.ps}{radec_secondary.ps}}
\vbox{\plottwo{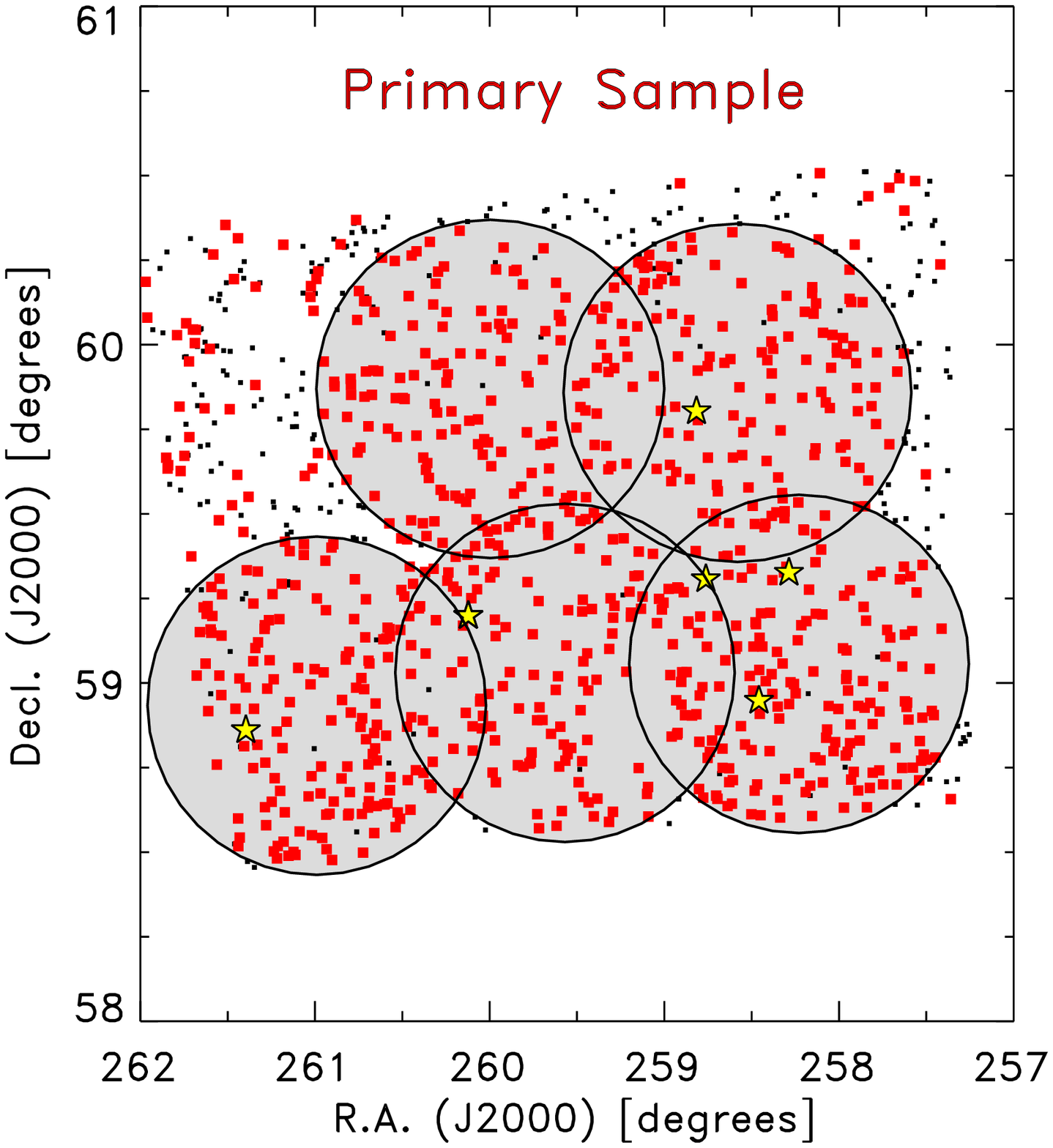}{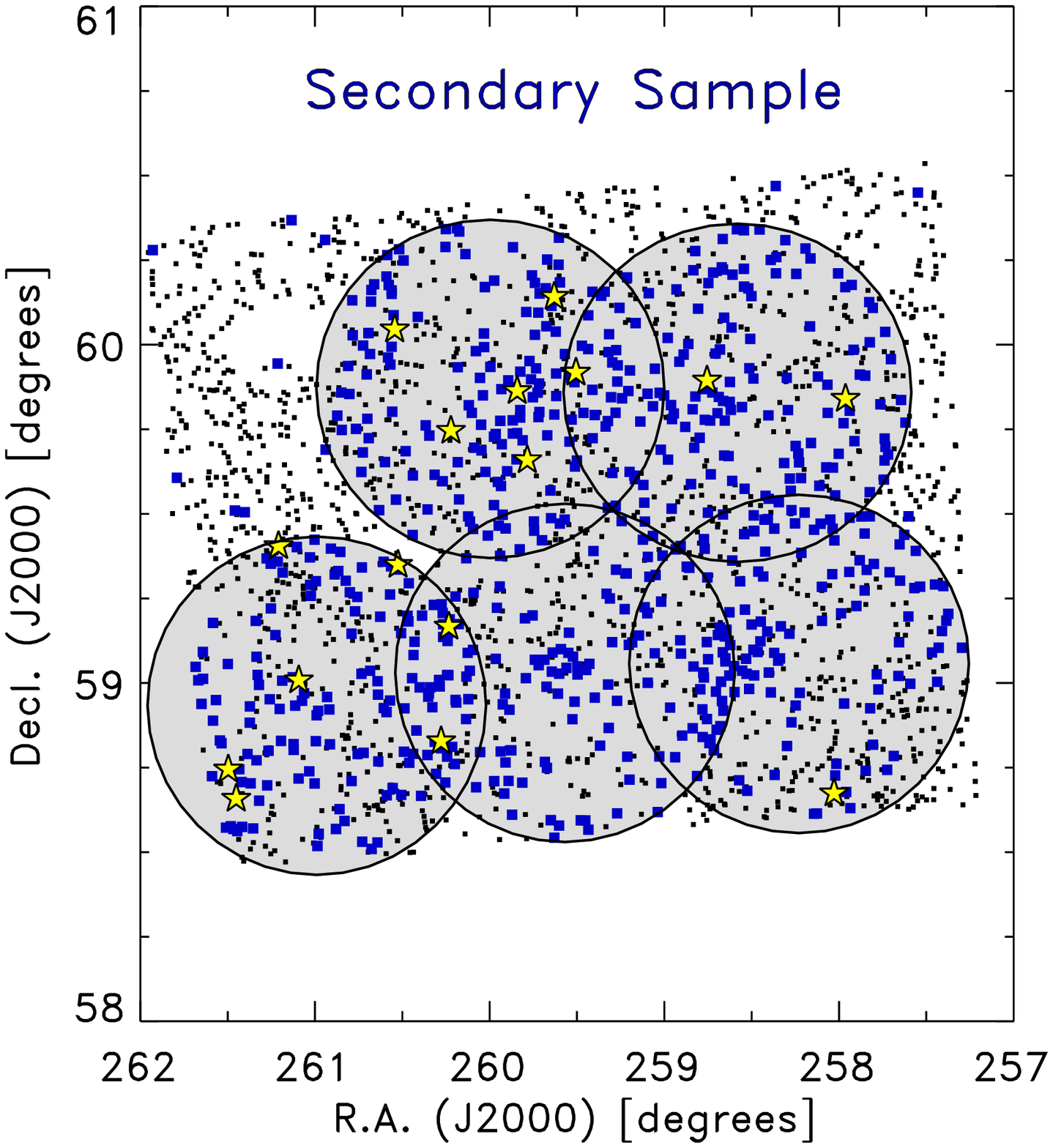}}
\epsscale{1.0}
\caption\figradec
\end{figure}
%TRUEMODE\else
%TRUEMODE\begin{figure*}
%TRUEMODE\epsscale{0.666}
%TRUEMODE\vbox{\plotone{radec.ps}}
%TRUEMODE\epsscale{1.15}
%TRUEMODE\vbox{\plottwo{radec_primary.ps}{radec_secondary.ps}}
%TRUEMODE\epsscale{1.0}
%TRUEMODE\caption\figradec
%TRUEMODE\end{figure*}
%TRUEMODE\fi

\section{Observations and Data Reduction}

\subsection{Observational Layout}

Hectospec is a 300--fiber spectrograph covering a 1 degree--diameter
field--of--view at the f/5 focus of the 6.5~m MMT \citep{fab05}.  It
began routine observing in 2004 April.  Each fiber aperture is
$1\farcs5$ diameter,  and the resulting spectra cover a wavelength
range of $\lambda=3500-9000$~\AA\ with 6~\AA\ FWHM resolution ($R\sim
1000$).

Figure~\ref{fig:radec} shows the \spitzer\ FLS layout including the
locations of the Hectospec fields, the 24~\micron\ parent sample, and
the 24~\micron\ sources with successfully measured redshifts.   Each
pointing provides some overlap with other pointings on the sky.  The
total area surveyed by the five Hectospec pointings is 3.3~deg$^2$,
correcting for the overlap between the fields.
 
We generated FLS fiber configurations for five different fields,
targeting sources from the parent sample defined in
\S~\ref{section:select}.   We named the configuration fields 131 to 135.  We
required that Hectospec place fibers on  4--7 spectrophotometric
F--type stars selected from SDSS, and we placed $\sim$30 fibers on
blank--sky locations to measure the sky brightness.   The remaining
fibers (typically $>$250) were placed on sources from the parent
sample, giving the highest priority to objects in the primary sample,
and lowest priority to objects in the tertiary sample.   We executed
our observations during 2004 July using an exposure time of 45 minutes
split into 3$\times$15~minute exposures.   Seeing was typically
sub-arcsecond.   Table~\ref{table:obs} summarizes the Hectospec field
centers and log for the observations, including the airmass at the
time of the observation.  In total, we targeted 1291 sources for
spectroscopy (605 primary targets, 632 secondary targets, 23 tertiary
targets, and 31 filler targets); not including sky or unused fibers,
nor calibration stars.

\subsection{Data Reduction}\label{section:redux}

We reduced the Hectospec data using
HSRED\footnote{http://mizar.as.arizona.edu/$\sim$rcool/hsred/}, an IDL
package developed by one of us (R.~Cool) for the reduction of data
from the Hectospec and Hectochelle  spectrographs
\citep{fab05,sze98}.  HSRED is based heavily on the reduction routines
developed for SDSS. Initially, cosmic rays are identified and removed
from the two--dimensional images using the IDL version of L.~A.\ Cosmic
\citep{vandok01} developed by J.~Bloom.   Dome flat observations are
used to identify the 300 fiber traces on the two--dimensional images.
These provide a high--frequency flat field and fringing correction for
the object spectra.  On nights when twilight sky spectra were
obtained, these images were used to derive a low--order correction to
the flat field vector for each fiber.   We refine the wavelength
solution for each observation, originally determined from HeNeAr
comparison spectra, using the locations of bright sky lines in each
object spectrum.   We further use the strengths of several sky
features to determine a small amplitude correction for each fiber to
remove relative transmission differences between fibers, which are not
fully removed by the flat field.  The local sky spectrum for each
observation is determined using the $\sim$30 dedicated sky fibers,
and is then subtracted from each object spectrum.

We flux--calibrate the spectra using simultaneous observations of the
F--stars on each Hectospec configuration.  The spectral class of each
F--star is determined by a comparison against a grid of \citet{kur93}
atmospheric models.  The ratio between the observed spectral shape of
the F--star and the best fit model provides the shape of the
sensitivity function of each fiber.  SDSS photometric observations of
the standard stars are used to normalize the absolute flux scale.
After each exposure is extracted, corrected for helio--centric motion
($\approx$6~km~s$^{-1}$, given the placement of the FLS in the \spitzer\
CVZ), and flux calibrated, we de--redden each spectrum according to
the Galactic dust maps of \citet{sch98}  using the \citet{odon94}
extinction curve.  Lastly, we coadd multiple exposures of a field to
obtain the final, flux--calibrated spectra.

\subsection{Problems with the Atmospheric Dispersion Corrector}
\label{section:adc}

Hectospec uses a counter--rotating atmospheric dispersion corrector
(ADC) to counteract the atmosphere--induced wavelength--dependent
aberration.   Unfortunately, during the first few months of Hectospec
operation (prior to 2004 October) the ADC was erroneously rotating in
the wrong direction.  As a result, our spectroscopic data suffer from
double the usual wavelength  aberration.  As guiding is in the visual,
light in the blue and red tend to miss the $1\farcs5$--diameter fiber
aperture, in an amount that depends on  the surface brightness profile
of the object.  Point sources suffer more than extended ones.

As discussed in \S~\ref{section:redux}, we attempted to
flux--calibrate the spectra by matching the spectrophotometric F--type
stars for each configuration to a grid of model atmospheres.   Because
of the atmospheric dispersion,  the flux--calibration correction in
the blue can be large, and for extended  sources, it is typically
overcorrected.   Even point sources need  not have the same
calibration due to slight differences in the  centroiding of the
target within the fiber.   Figure~\ref{fig:adc} shows examples of
several flux--calibrated spectra from our data from three different
configurations, taken over the range of airmass 1.16--1.35 (see
table~\ref{table:obs}).   There is no way to correct for the problem
with the ADC, so we do  not recommend these data for applications
requiring accurately flux-calibrated spectra over long wavelength
baselines.  However, the flux calibration in the mid-optical, e.g.,
5000--7000~\AA, should be less affected by the ADC problem and may be
useful.  Line indices in this wavelength range should
be robust, as should flux ratios of lines of comparable wavelength
(e.g., [\ion{O}{3}$\lambda5007$]/\hb,
\ha/[\ion{N}{2}$\lambda6585$]).  Moreover, the ADC problems do not
strongly effect the primary focus of measuring redshifts from the
spectra.   
%TRUEMODE\ifsubmode
\begin{figure}
%\plotone{samplespec.ps}
\plotone{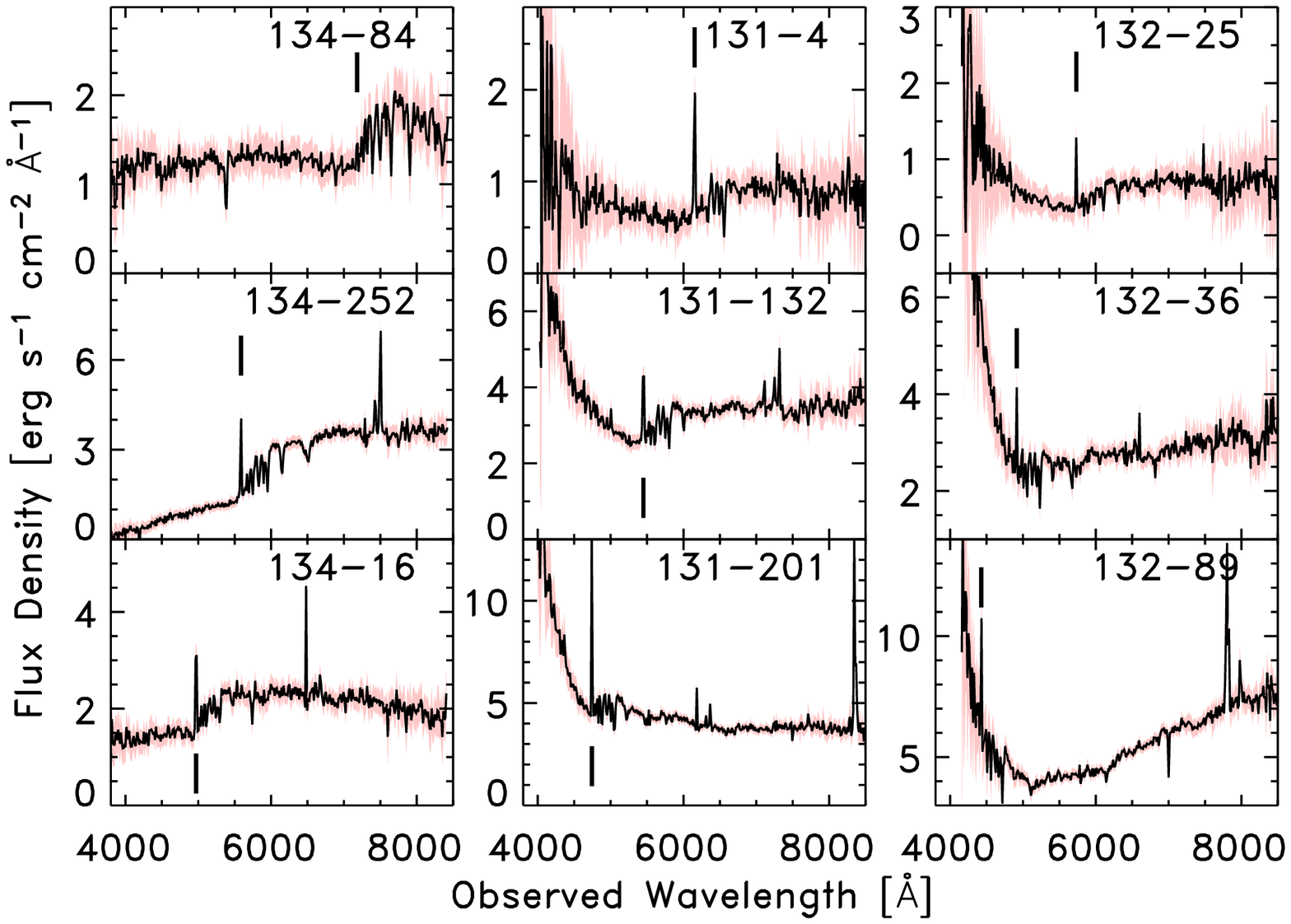}
\caption\figcapsamplespec
\end{figure}
%TRUEMODE\else
%TRUEMODE\begin{figure*}
%TRUEMODE\plotone{samplespec.ps}
%TRUEMODE\caption\figcapsamplespec
%TRUEMODE\end{figure*}
%TRUEMODE\fi

\subsection{Redshift Measurements}

We measured redshifts from the MMT/Hectospec spectra
using first an automated pipeline analysis, converted from the SDSS
pipeline.  We subsequently inspected each spectrum to check the
automated measurement.  The observing program was highly efficient:
of the 1291 science targets for spectroscopy, we measured redshifts
successfully for 1270 (a 98\% success rate; including both those
classified from the automated pipeline and those visually identified). 

For our automated measurements, we used a version of the SDSS pipeline
{\tt specBS} (D.~Schlegel \etal\ 2006;
http://spectro.astro.princeton.edu) adapted for MMT/Hectospec for the
AGN and Galaxy Evolution Survey (AGES) of the NDWFS in the
constellation Bo\"otes \citep{koc06} and included in the HSRED
package (see \S~\ref{section:redux}).   {\tt SpecBS} uses $\chi^2$
minimization to compare each spectrum to various model spaces, each a
linear combination of eigentemplates at different redshifts.  The
minimum $\chi^2$ yields the object's spectral classification (star,
galaxy, QSO, and any sub-classification) as well as the redshift. The
classifications are included in table~\ref{table:mmtcatalog}.   The
pipeline also provides a warning flag for the redshift (see
\S~\ref{section:mmtspectra}).

We then visually inspected the spectra to verify the
accuracy of the pipeline--derived redshifts.  The automated
classification scheme had a high success rate: of the 1291 scientific
objects we targeted with fibers, we found that the automated
classification algorithm returned faulty or dubious redshifts in only
74 cases (5\%).  Most of these cases occur in observations at high
airmass, where the spectra typically have excessive
blue--light flux (see \S~\ref{section:adc}).  Roughly 50\% (36/74) of
objects with erroneous redshifts are from field 132 (airmass 1.35).
The remaining erroneous redshifts are split fairly evenly between the
three configurations taken at moderate airmass (131, 133, and 135).
We found only five erroneous redshifts in field 134, with airmass 1.16.

For the majority of spectra with erroneous redshifts (53/74), we
identified an alternative redshift by visual inspection.  We then
remeasured a redshift by fitting a Gaussian to the centroid of strong
features in the spectra (normally [\ion{O}{2}] or [\ha], depending on
the redshift coverage, and in some cases the Ca H+K doublet) and
corrected these to the heliocentric frame.  We inserted these
remeasured redshifts and uncertainties into the MMT/Hectospec catalog,
and set \texttt{ZVISUALFIXFLAG}=1 for these.   For these objects, we
visually identified the spectroscopic classification as either
``Galaxy'', ``QSO'', or ``STAR''. Note that like SDSS, ``QSO'' refers
to broad--line QSOs; objects identified as a ``Galaxy'' include also
sources that may have putative narrow--line AGN based on broad
[\ion{O}{3}], \ha, or line ratios.  We also updated the spectroscopic
subclassification to ``manual'' for objects where we remeasured the
redshift.
%These remeasured redshifts have offsets
%from those derived from the pipeline, due to differences in the
%air--to--vacuum wavelength shift. (Note that the correction to the
%heliocentric frame is $\approx$6~km~s$^{-1}$ given the location of the
%FLS in the \spitzer\ CVZ.)   To correct for this effect and for any
%other systematics between our remeasured redshifts and the automated
%ones, we rederived the redshifts of 50 objects with high
%signal--to--noise spectra and robust pipeline--measured redshifts
%selected evenly across the five configurations.  The mean offset
%between the remeasured and pipeline redshifts is $81$~km~s$^{-1}$
%(consistent with known air--to--vacuum offsets, see Morton 1991),
%which we applied to the objects with remeasured redshifts.  

Table~\ref{table:mmtcatalog} lists Hectospec spectroscopic results,
including the automated redshifts, redshift uncertainty (columns
10--11), and  spectroscopic classification and any subclassification
(columns 8 and 9).   These entries include objects where we manually
measured the redshift, and modified the spectroscopic classification.
Objects with visual--flag values of 1 (column 14) correspond to those
we judged to have secure redshifts.   Although we trust all objects
with visual--flag values of 1, for completeness we include the
redshift warning flag ({\tt ZWARNING}; column 12), and reduced
$\chi^2$ of the redshift fit (column 13) from the automated
measurement.  We consider objects with visual--flag values of 0 to
have untrustworthy redshifts, although we have left the
pipeline--derived redshift in the catalog for completeness.   Column 4
gives the field number from table~\ref{table:obs} of the observation,
column 5 gives the Hectospec number of the fiber used to target the
object, and column 6 gives the reduction number.  The  ID numbers in
table~\ref{table:mmtcatalog} column 1 correspond to the ID numbers in
column 1 of table~\ref{table:catalog}.  Most of the objects with
visual--flag values of 1 have good automated redshift measurements
(with {\tt ZWARNING}=0; column 12), but 63 of these objects have {\tt
ZWARNING}$>$0.  Many of the {\tt ZWARNING} $>$0 objects have high
signal--to--noise spectra, although some (31/63) are ones for which we
identified an alternative redshift.  Objects with fix--flag values of
1 (\texttt{ZVISUALFIXFLAG}; column~15) correspond to objects for which we visually remeasured
an alternative redshift as described above.

As part of the electronic edition of this publication, we present the
full photometry and redshift catalog in three separate files.  The pertinent information
in these three files is given in
tables~\ref{table:catalog}--\ref{table:mmtcatalog}.   The photometry
file contains astrometry and photometry from SDSS and MIPS for objects
in a parent sample that includes all MIPS sources in the FLS with SDSS
matches at $\isdss<21$~mag, 37 calibration stars, and 31 other targets
that we observed on these configurations as filler targets (see
below).  Two other files contain the spectroscopic results from the
MMT and from the SDSS, respectively.  These three files are row
synchronous; each file contains 7226 objects, including the 5698
objects in the primary, secondary, and tertiary samples.  The ID
numbers in column 1 of
tables~\ref{table:catalog}--\ref{table:mmtcatalog} correspond to the
row of the object in each of the three catalog files.

\section{Spectroscopic Completeness}

As discussed in \S~\ref{section:select}, our targeting criteria select
$\approx$60\% of all 24~\micron\ sources with $f_\nu(24\micron) \geq
1$~mJy, and $\approx$30\% of those sources with 0.3~mJy $\leq
f_\nu(24\micron) <$~1~mJy.   We did not observe MIPS 24~\micron\
sources with optical magnitudes fainter than $\isdss > 21$
and 20.5~mag for the primary and secondary samples, respectively.
\begin{figure}
\plotone{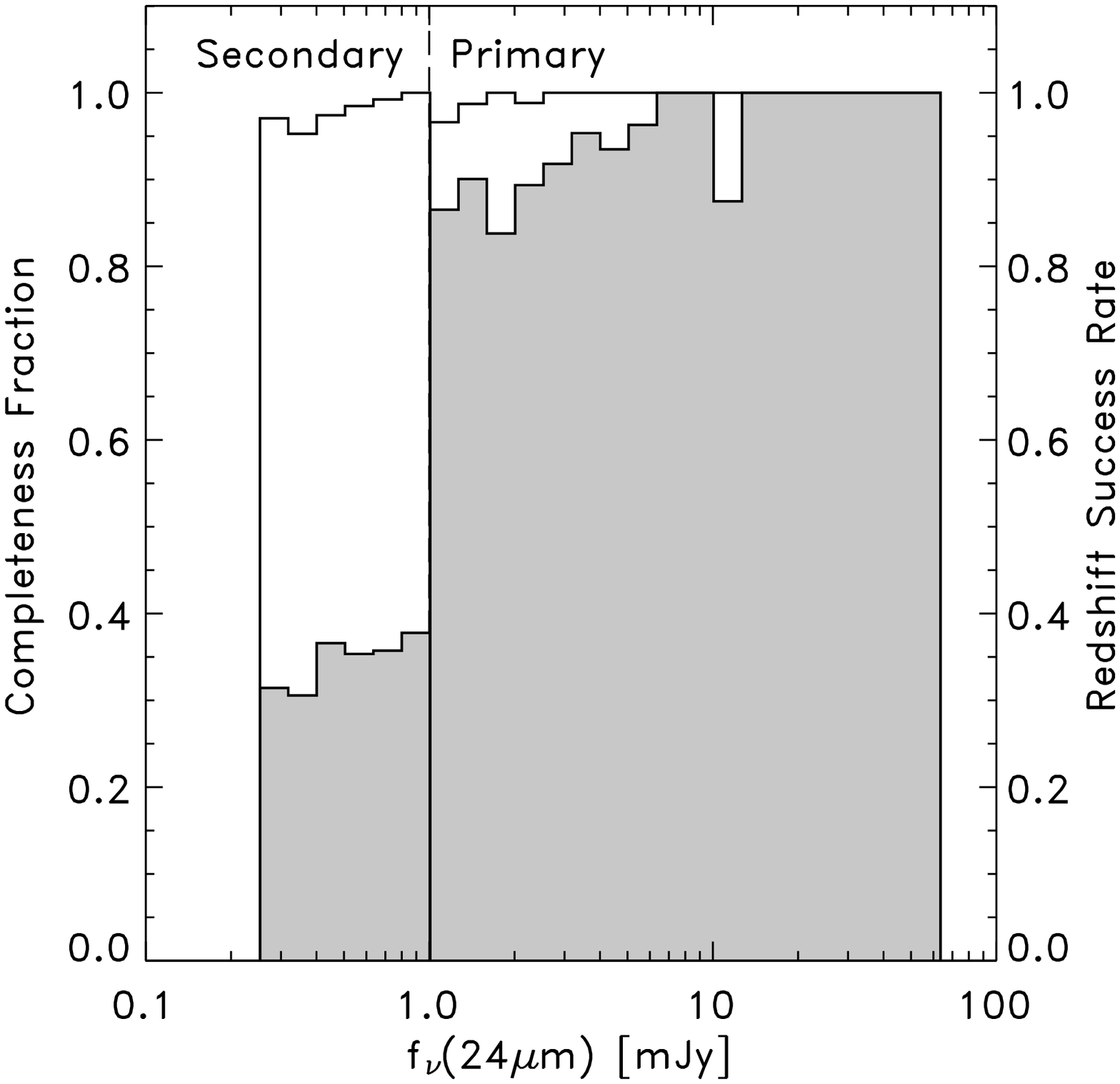}
\caption\figcapcomp
\end{figure}

The total spectroscopic success rate is high.   We define the
redshift--success rate as the ratio of the number of Hectospec targets
with measured redshifts to the number of Hectospec targets observed,
which we show in figure~\ref{fig:comp}.   The redshift success rate
for both the primary and secondary samples is 98--99\% (599/605
targeted primary sources, and 619/632 targeted secondary sources have
successfully measured redshifts).

We compute the overall spectroscopic completeness as the ratio of
targets with successfully measured redshifts (including both the
MMT/Hectospec and SDSS spectroscopy) to the subsample of targets in
the parent sample that fall within our Hectospec fields, i.e.,
including only those objects that were targets or potential targets
for our configurations.   Figure~\ref{fig:comp} shows the
spectroscopic completeness as a function of 24~\micron\ flux density.
Objects without measured redshifts are counted once for each Hectospec
field in which they occur (see figure~\ref{fig:radec}).
%The
%secondary and tertiary samples have lower priorities in the Hectospec
%configurations compared to the primary sample.  Combined with the
%larger surface density of these targets (see figure~\ref{fig:number}),
%this causes the lower spectroscopic completeness.
There are 820 objects in the primary sample falling in the region
observed with Hectospec.   Of these, 739 have redshifts, with 140 from
the SDSS and 599 from our Hectospec program.  This is a completeness
of 90\%.   The 10\% that were not observed tend to  be near the outer
edge of the observed region (see fig~\ref{fig:radec}).  Hectospec
fibers extend nearly radial from the edge of the field of view, and
near the edge only one fiber can serve a target, leading to
diminishing completeness.  Pairs closer than the minimum Hectospec
fiber approach of 20\arcsec\ will also have only one object observed,
unless one of the objects was bright enough to be observed by SDSS or
unless the pair falls in the overlap region of two MMT pointings.
Aside from these minor issues, the primary sample completeness does
not have any significant spatial biases.

Of the 1771 objects in the secondary sample falling in the observed
region, 662 have redshifts, with 43 from the SDSS and 619 from our
Hectospec program (37\% completeness).  However, we stress that this
completeness is highly spatially biased.  One can clearly see
large-scale structure in the sample that is not being tracked by the
subset with spectra.   We have not modeled this incompleteness yet,
and one should be  cautious about using the secondary sample for
statistical applications  that would be affected by a sampling that
depends on the density of  targets on the sky.

The tertiary sample is very sparsely covered ($< 3$\% complete) and
should not be used for  statistical work.

\section{The MMT/Hectospec Spectra}\label{section:mmtspectra} 

As part of this publication, we make available the fully reduced
individual MMT/Hectospec spectra.   These data are included as
multi-extension FITS (MEF) files as part of the electronic edition of
this publication.  They are also available through the NASA/IPAC
Infrared Science Archive (IRSA; see url in
footnote~\ref{footnote:url}).  The format of the MEF files for the
spectra is similar to those from the SDSS pipeline
(http://spectro.astro.princeton.edu).  There are 10 Header Data Unit
(HDU) extensions in each MEF.  Extensions 1--5 contain information for
the spectra; these have dimension of $N_\mathrm{pixel} \times
N_\mathrm{fiber}$, where $N_\mathrm{pixel} = 4608$ is the number of
pixels in the spectrum from each fiber and $N_\mathrm{fiber} = 300$ is
the number of fibers.  Extension 6 contains targeting information for
each fiber.  Extensions 7--10 contain information on the flat field and sky
spectrum from the data reduction.  They are not useful for most
applications.   The HDUs are:
\begin{itemize}
\item HDU 1: Wavelength Solution in \AA ngstroms in vacuum;
\item HDU 2:  Flux density of the spectrum in units of $10^{-17}$~erg s$^{-1}$
cm$^{-2}$ \AA$^{-1}$;
\item HDU 3:  Inverse variance ($\sigma^{-2}$) for the flux
density;
\item HDU 4: Mask of warning flags in each pixel combined
between all exposures with logical AND;
\item HDU 5: Mask of warning flags in each pixel combined
between all exposures with logical OR;
\item HDU 6: Structure containing targeting information for each fiber, it contains:
\begin{itemize}
	\item \texttt{OBJTYPE}: Type of fiber assignment (target, sky,
	standard star, or rejected); \item \texttt{RA}: Right
	Ascension of fiber (deg; J2000); \item \texttt{DEC}:
	Declination of fiber (deg; J2000); \item \texttt{FIBERID}:
	Maps aperture to physical fiber on Hectospec; \item
	\texttt{RMAG}: SDSS \rsdss--band magnitude of source; \item
	\texttt{RAPMAG}: Estimate of the \rsdss--band magnitude
	measured through an aperture with $1\farcs5$ diameter; \item
	\texttt{ICODE}: Integer targeting flags for the observation;
	\item \texttt{RCODE}: Not used; \item \texttt{BCODE}: Not
	used; \item \texttt{MAG}: SDSS magnitudes for objects with
\texttt{OBJTYPE} equal to standard star;
	\item \texttt{XFOCAL}: $x$--coordinate of fiber on Hectospec
focal plane;
	\item \texttt{YFOCAL}: $y$--coordinate of fiber on Hectospec
focal plane; 
	\item \texttt{FRAMES}: Bookkeeping value from reductions;
	\item \texttt{EXPID}: Name of highest--quality exposure;
	\item \texttt{TSOBJID}: Not used;
	\item \texttt{TSOBJ\_MAG}: Bookkeeping values used in spectra
flux calibration;
	\item \texttt{EBV\_SFD}: Galactic extinction, $E(B-V)$, for
fiber derived from \citet{sch98}.
	\end{itemize}
\item HDU 7: Structure containing B--Spline parameters for the sky
spectrum derived for fibers 1--150;
\item HDU 8: Array of dimension $N_\mathrm{pixel}\times 150$
containing the auxiliary flat--field correction derived for fibers
1--150 based on strength of sky lines as part of the reduction;
\item HDU 9: Same as HDU~6, but for fibers 151--300;
\item HDU 10: Same as HDU~8, but for fibers 151--300.
\end{itemize}
We remind the reader here that the flux calibration
suffers from problems arising from the ADC (see \S~\ref{section:adc});
caution should be used when using these spectra for applications
required accurate flux calibration.  The field and fiber value for
each object in table~\ref{table:mmtcatalog} correspond to the array
for each fiber in MEF for each field. 

The data files also contain spectra of 57 objects not part of the 
MIPS sample.  Of these, 26 are calibration stars; the other 31 are
from a filler sample of $z\sim0.5$ red galaxies and quasars, which 
are far too sparse to be statistically useful.

\section{Discussion and Summary}

We have obtained 1296 redshifts for \spitzer\ 24~\micron--selected
sources in the \spitzer\ FLS using the Hectospec fiber spectrograph on
the MMT.  Our observing program was highly efficient ($\simeq$98--99\%
redshift success rate).  It is 90\% complete for $f_\nu(24\micron) \geq
1$~mJy and $\isdss \leq 21$~mag, and is 37\% complete for 0.3~mJy $<
f_\nu(24~\micron) <$ 1~mJy and $\isdss \leq 20.5$~mag. As part of this
publication we provide catalogs for the full parent sample, and the
SDSS and  MMT redshift catalogs.  We also publish our reduced,
flux--calibrated spectroscopic Hectospec data.

Our spectroscopic survey of the \spitzer\ FLS identifies galaxies and
QSOs over a large redshift range.  Figure~\ref{fig:zhist} shows the
redshift distribution for the 24~\micron\ sources from the primary,
secondary, and tertiary samples in the FLS, including 1246 sources
identified as galaxies and QSOs from our MMT/Hectospec survey, and 280
additional galaxies and QSOs with redshifts from SDSS.  Our
Hectospec survey of 24~\micron\ sources identifies galaxies to
$z\leq 0.98$ and QSOs to $z\leq 3.6$.   In figure~\ref{fig:maxz}, we
show the MMT/Hectospec spectra of the highest redshift objects
spectroscopically classified as a galaxy and QSO.  

%TRUEMODE\ifsubmode
\begin{figure}
\epsscale{0.9}
\plotone{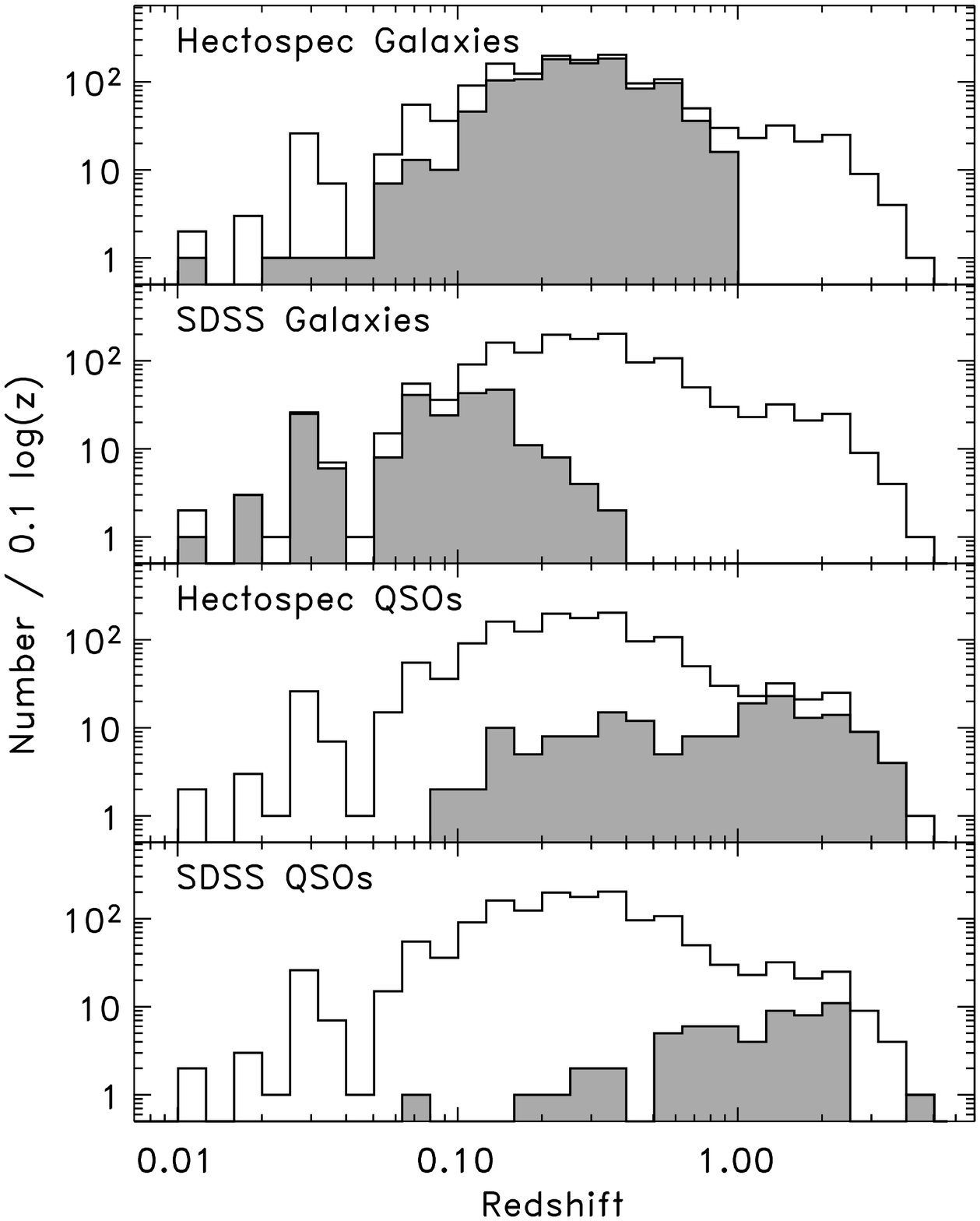}
%\plotone{zdist.ps}
\epsscale{1.0}
\caption\figcapzhist
\end{figure}
%TRUEMODE\else
%TRUEMODE\begin{figure*}
%TRUEMODE\epsscale{0.9}
%TRUEMODE\plotone{zdist.ps}
%TRUEMODE\epsscale{1.0}
%TRUEMODE\caption\figcapzhist
%TRUEMODE\end{figure*}

%TRUEMODE\fi
%TRUEMODE\ifsubmode
\begin{figure}
\epsscale{0.75}
\vbox{\plotone{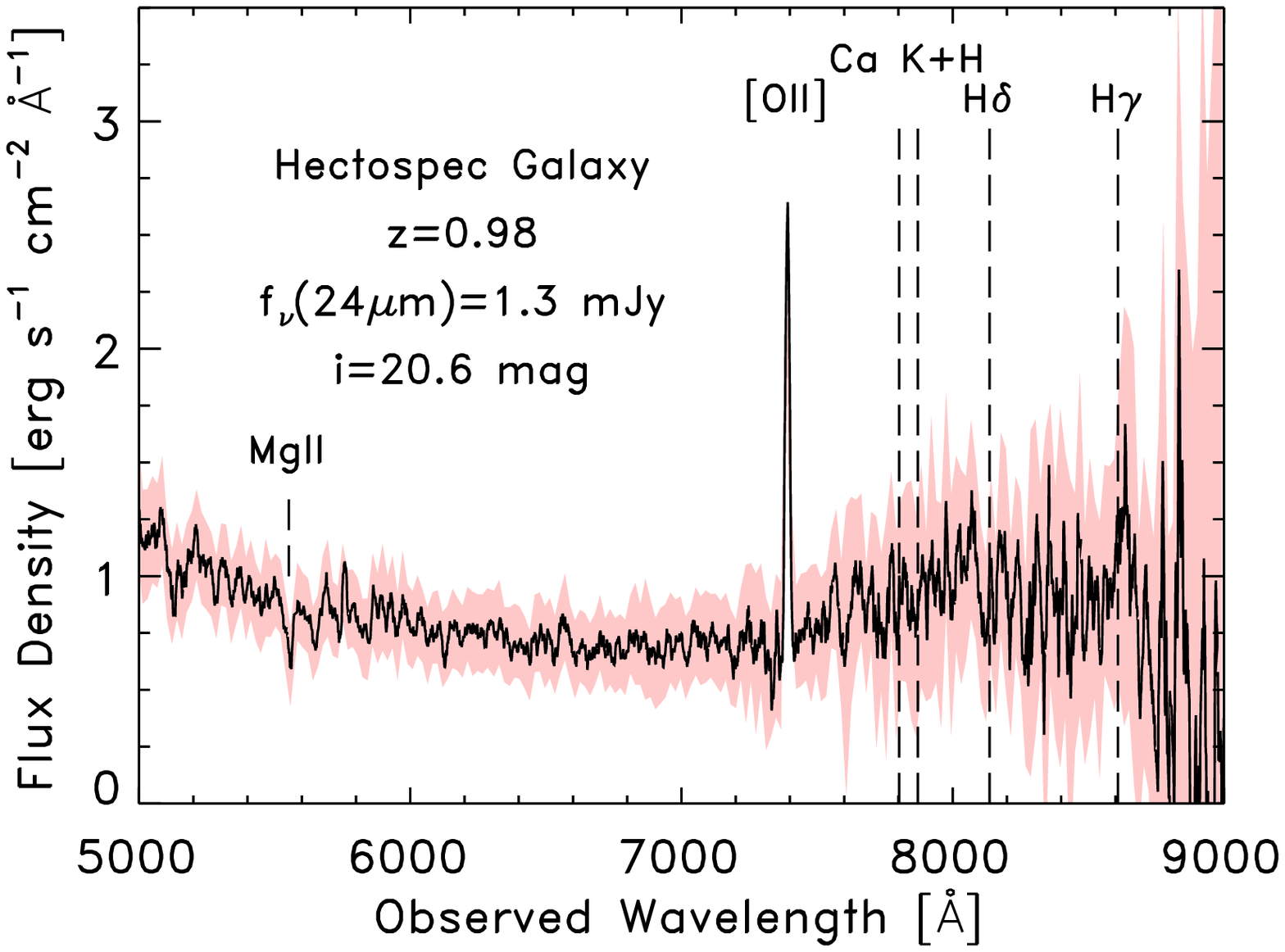}}
%\vbox{\plotone{maxz_gal.ps}}
\vspace{0.15in}
\vbox{\plotone{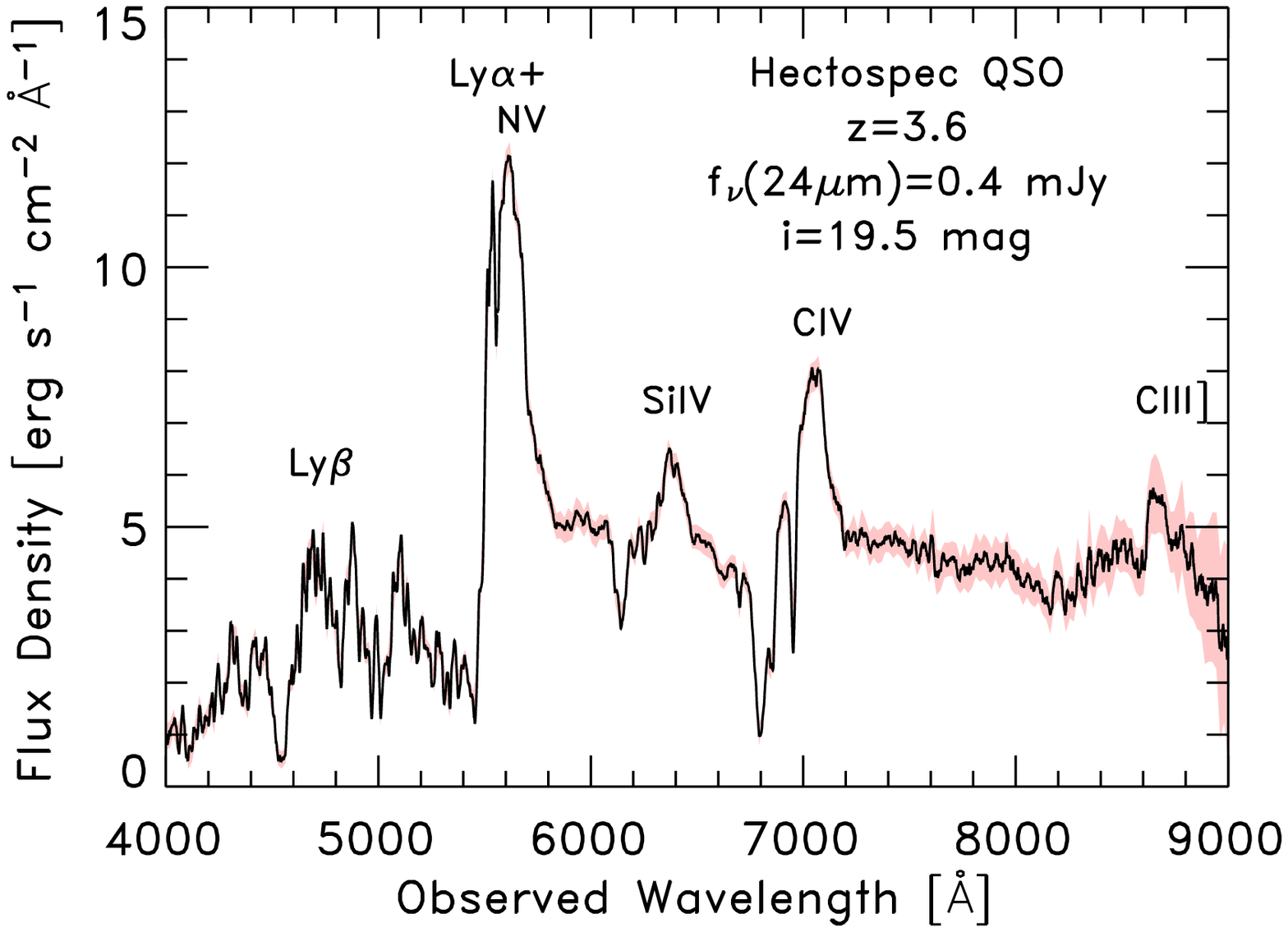}}
%\vbox{\plotone{maxz_qso.ps}}
\caption\figcapmaxz
\end{figure}
%TRUEMODE\else
%TRUEMODE\begin{figure*}
%TRUEMODE\plottwo{maxz_gal.ps}{maxz_qso.ps}
%TRUEMODE\epsscale{1.0}
%TRUEMODE\caption\figcapmaxz
%TRUEMODE\end{figure*}
%TRUEMODE\fi

The redshift distribution for galaxies spans  $0.01 \lsim z_\mathrm{gal}
\lsim 1$, and the distribution for QSOs extends to higher redshift
$0.1 \lsim z_\mathrm{QSO} \lsim 5$.   The mean redshifts for galaxies
and QSOs (including the Hectospec and SDSS data) are $\langle
z_\mathrm{gal} \rangle = 0.28$ and $\langle z_\mathrm{QSO} \rangle =
1.1$, with standard deviations of $\sigma_\mathrm{gal} = 0.17$ and
$\sigma_\mathrm{QSO} = 0.84$.   For the primary sample, the mean
redshifts for galaxies and QSOs are $\langle z_\mathrm{gal} \rangle =
0.24$ and $\langle z_\mathrm{QSO}
\rangle = 1.0$, with standard deviations of $\sigma_\mathrm{gal} =
0.17$ and $\sigma_\mathrm{QSO} = 0.77$.   Similarly, for the secondary
sample, the mean redshifts for galaxies and QSOs are $\langle
z_\mathrm{gal} \rangle = 0.31$ and $\langle z_\mathrm{QSO} \rangle =
1.5$, with standard deviations of $\sigma_\mathrm{gal} = 0.16$ and
$\sigma_\mathrm{QSO} = 0.94$.

The photometric and redshift catalogs, and reduced spectra are
available with the electronic edition of this publication.  They are
also available through the NASA/IPAC Infrared Science Archive (IRSA;
see footnote~\ref{footnote:url}). We are
currently pursuing further redshift surveys in \spitzer\ fields,
building the datasets needed to understand the IR--active galaxy
population.  

%The redshift catalogs for the \spitzer\ FLS MIPS 24~\micron\ sources
%have a large range of applications.  We are currently using these data
%combined with similar datasets in other fields to study the 
%properties and evolution of the 24~\micron\ sources to $z$$\sim$1.
%These data have also been used to construct revised spectral energy
%distributions and bolometric corrections for QSOs over a large range
%of wavelength and redshift \citep{ric06}, and we are studying the
%mid--Infrared properties with \spitzer\ spectroscopy for a subset of
%galaxies drawn from our catalog.  

\acknowledgments

We wish to thank Tom Soifer and the SSC staff for initiating the FLS
project, and for their hard work processing the \spitzer\ data.  We
thank Michael Cushing, John Moustakas, and J.~Davy Kirkpatrick for
useful conversations and feedback, and we thank Lisa
Storrie--Lombardi, Anastasia Alexov, Bruce Berriman, and J.~Davy
Kirkpatrick for their assistance in providing the data
products through IRSA.  We are grateful to the referee, Bahram
Mobasher, for useful comments and suggestions.   We thank the
Hectospec instrument and queue--mode scientists and the MMT operators
for their excellent assistance with the observations.   Observations
reported here were obtained at the MMT Observatory, a joint facility
of the University of Arizona and the Smithsonian Institution.  This
work is also based in part on archival data obtained with the Spitzer
Space Telescope, which is operated by the Jet Propulsion Laboratory
(JPL), California Institute of Technology (Caltech) under a contract
with NASA. Support for this work was provided by awards issued by
JPL/Caltech, and by NASA through the Spitzer Space Telescope
Fellowship Program, through a contract issued by JPL, Caltech under a
contract with NASA.

This paper made use of data from the SDSS. Funding for the SDSS has
been provided by the Alfred P. Sloan Foundation, the Participating
Institutions, the National Aeronautics and Space Administration, the
National Science Foundation, the U.S. Department of Energy, the
Japanese Monbukagakusho, and the Max Planck Society. The SDSS Web site
is http://www.sdss.org/.  The SDSS is managed by the Astrophysical
Research Consortium (ARC) for the Participating Institutions. The
Participating Institutions are The University of Chicago, Fermilab,
the Institute for Advanced Study, the Japan Participation Group, The
Johns Hopkins University, the Korean Scientist Group, Los Alamos
National Laboratory, the Max-Planck-Institute for Astronomy (MPIA),
the Max-Planck-Institute for Astrophysics (MPA), New Mexico State
University, University of Pittsburgh, University of Portsmouth,
Princeton University, the United States Naval Observatory, and the
University of Washington.

%\begin{thebibliography}
%\include{refs}

%\end{thebibliography}

\begin{deluxetable}{lc}
\tableone
\end{deluxetable}

\begin{deluxetable}{lccccc}
\tabletwo
\end{deluxetable}

%TRUEMODE\ifincludetable
\begin{deluxetable}{lccccccccccccc}
\tabletypesize{\scriptsize}
\tablecolumns{14}
\rotate
\tablewidth{0pt}
\tablecaption{SDSS--Matched 24~\micron\ Catalog\label{table:catalog}}
\tablehead{ \colhead{} & 
\colhead{R.A.} & 
\colhead{Decl.} & 
\colhead{$i$} & 
\colhead{$u$$-$$g$} & 
\colhead{$g$$-$$r$} & 
\colhead{$r$$-$$i$} & 
\colhead{$i$$-$$z$} & 
\colhead{$A(r)$} & 
\colhead{Targ.} & 
\colhead{$\log f_\nu(24\micron)$} & 
\colhead{} & 
\colhead{R.A.\ (24\micron)} & 
\colhead{Decl.\ (24\micron)} \\
\colhead{ID} & 
\colhead{(J2000)} & 
\colhead{(J2000)} & 
\colhead{(mag)} & 
\colhead{(mag)} & 
\colhead{(mag)} & 
\colhead{(mag)} & 
\colhead{(mag)} & 
\colhead{(mag)} & 
\colhead{Flag} & 
\colhead{(mJy)} & 
\colhead{SNR$_{24}$} & 
\colhead{(J2000)} & 
\colhead{(J2000)} \\ 
\colhead{(1)} & 
\colhead{(2)} & 
\colhead{(3)} & 
\colhead{(4)} & 
\colhead{(5)} & 
\colhead{(6)} & 
\colhead{(7)} & 
\colhead{(8)} & 
\colhead{(9)} & 
\colhead{(10)} & 
\colhead{(11)} & 
\colhead{(12)} & 
\colhead{(13)} & 
\colhead{(14)}}
\startdata
    1 &  17\phn17\phn50.52 & +60\phn26\phn48.1 & 18.01 & \phs1.30 & \phs0.50 & \phs0.15 & \phs0.03 & 0.066 & 8 & $-$1.33 & \phn\phn  0.98 &  17\phn17\phn50.40 & +60\phn26\phn47.9 \\
    2 &  17\phn17\phn55.48 & +60\phn26\phn39.3 & 17.59 & \phs1.92 & \phs1.28 & \phs0.50 & \phs0.32 & 0.066 & 4 & $-$0.15 & \phn\phn  3.51 &  17\phn17\phn55.53 & +60\phn26\phn39.6 \\
    3 &  17\phn18\phn02.38 & +60\phn27\phn20.8 & 18.60 & \phs1.09 & \phs1.36 & \phs0.51 & \phs0.27 & 0.067 & 8 & $-$0.74 & \phn\phn  2.29 &  17\phn18\phn02.35 & +60\phn27\phn20.7 \\
    4 &  17\phn17\phn53.86 & +60\phn26\phn20.2 & 20.36 & \phs2.54 & \phs0.91 & \phs0.40 & \phs0.46 & 0.065 & 4 & $-$0.40 & \phn\phn  2.58 &  17\phn17\phn54.02 & +60\phn26\phn21.3 \\
    5 &  17\phn18\phn11.89 & +60\phn26\phn49.3 & 20.19 & \phs1.71 & \phs1.50 & \phs0.39 & \phs0.52 & 0.066 & 4 & $-$0.39 & \phn\phn  2.58 &  17\phn18\phn11.89 & +60\phn26\phn49.3 \\
    6 &  17\phn17\phn30.78 & +60\phn26\phn37.6 & 20.84 & $-$0.25 & \phs1.59 & \phs0.48 & \phs0.28 & 0.066 & 0 & $-$0.87 & \phn\phn  1.46 &  17\phn17\phn30.95 & +60\phn26\phn36.6 \\
    7 &  17\phn18\phn11.34 & +60\phn27\phn06.9 & 20.53 & \phs1.31 & \phs1.53 & \phs0.65 & \phs0.69 & 0.066 & 0 & $-$0.57 & \phn\phn  2.71 &  17\phn18\phn11.33 & +60\phn27\phn06.4 \\
    8 &  17\phn18\phn21.07 & +60\phn27\phn29.3 & 20.87 & \phs1.81 & \phs1.67 & \phs0.41 & \phs0.24 & 0.066 & 0 & $-$1.37 & \phn\phn  1.08 &  17\phn18\phn21.08 & +60\phn27\phn27.5 \\
    9 &  17\phn18\phn37.32 & +60\phn25\phn24.6 & 19.92 & \phs3.51 & \phs0.98 & \phs0.41 & \phs0.28 & 0.063 & 8 & $-$1.05 & \phn\phn  1.25 &  17\phn18\phn37.49 & +60\phn25\phn24.2 \\
   10 &  17\phn17\phn50.77 & +60\phn17\phn01.2 & 16.02 & \phs1.95 & \phs0.90 & \phs0.38 & \phs0.31 & 0.057 & 8 & $-$1.22 & \phn\phn  0.94 &  17\phn17\phn50.86 & +60\phn17\phn01.9 \\
   11 &  17\phn18\phn20.37 & +60\phn26\phn20.8 & 20.65 & \phs0.75 & \phs1.02 & \phs0.49 & \phs0.43 & 0.065 & 0 & $-$0.34 & \phn\phn  3.10 &  17\phn18\phn20.41 & +60\phn26\phn21.1 \\
   12 &  17\phn18\phn44.49 & +60\phn23\phn42.3 & 18.65 & \phs0.74 & \phs0.95 & \phs0.35 & \phs0.36 & 0.062 & 2 & \phs0.62 & \phn 11.85 &  17\phn18\phn44.56 & +60\phn23\phn41.8 \\
   13 &  17\phn18\phn31.60 & +60\phn21\phn00.7 & 17.29 & \phs1.72 & \phs1.08 & \phs0.52 & \phs0.41 & 0.059 & 2 & \phs0.04 & \phn\phn  5.31 &  17\phn18\phn31.63 & +60\phn21\phn00.8 \\
   14 &  17\phn18\phn33.48 & +60\phn26\phn16.9 & 20.96 & \phs1.42 & \phs1.31 & \phs0.44 & \phs0.24 & 0.065 & 0 & $-$0.56 & \phn\phn  2.47 &  17\phn18\phn33.52 & +60\phn26\phn15.6 \\
   15 &  17\phn17\phn41.68 & +60\phn25\phn14.3 & 17.98 & \phs1.18 & \phs0.52 & \phs0.30 & \phs0.12 & 0.064 & 8 & $-$0.69 & \phn\phn  2.06 &  17\phn17\phn41.63 & +60\phn25\phn13.0 \\
   16 &  17\phn17\phn34.21 & +60\phn23\phn50.7 & 19.01 & \phs0.27 & \phs0.49 & \phs0.35 & \phs0.27 & 0.062 & 2 & \phs0.20 & \phn\phn  6.64 &  17\phn17\phn34.23 & +60\phn23\phn52.0 \\
   17 &  17\phn17\phn50.54 & +60\phn23\phn29.5 & 19.15 & \phs0.61 & \phs0.66 & \phs0.30 & \phs0.04 & 0.060 & 4 & $-$0.34 & \phn\phn  2.83 &  17\phn17\phn50.67 & +60\phn23\phn29.3 \\
   18 &  17\phn18\phn41.06 & +60\phn23\phn59.1 & 19.54 & \phs0.24 & \phs0.22 & \phs0.35 & \phs0.10 & 0.062 & 2 & \phs0.02 & \phn\phn  5.47 &  17\phn18\phn41.09 & +60\phn23\phn58.5 \\
   19 &  17\phn18\phn47.92 & +60\phn24\phn09.4 & 18.73 & \phs1.28 & \phs1.16 & \phs0.34 & \phs0.41 & 0.062 & 4 & $-$0.10 & \phn\phn  4.31 &  17\phn18\phn47.94 & +60\phn24\phn09.7 \\
   20 &  17\phn18\phn47.42 & +60\phn23\phn30.9 & 19.55 & \phs0.35 & \phs1.13 & \phs0.38 & \phs0.30 & 0.062 & 4 & $-$0.25 & \phn\phn  3.04 &  17\phn18\phn47.62 & +60\phn23\phn30.4 \\
\enddata
\tablecomments{(1) Row synchronous ID; (2) SDSS right ascension in units of hours, minutes, and seconds; (3) SDSS declination in units of degrees, arcminutes, and arcseconds; (4) SDSS combined--model magnitude (dereddened); (5--8) SDSS combined--model colors (dereddened); (9) Extinction used to deredden magnitudes in (4--8); (10) Hectospec Target Flag (see text); (11) Logarithm of MIPS 24~\micron\ flux density; (12) MIPS 24~\micron\ signal--to--noise ratio, $\equiv f_\nu(24\micron)/\delta f_\nu(24\micron)$; note that 24~\micron\ flux--density errors do not include correlated pixel noise and are thus underestimated (see text); (13) right ascension of MIPS 24~\micron\ source in units of hours, minutes, and seconds; (14) declinations of MIPS 24~\micron\ source in units of degrees, arcminutes, and arcseconds.  Table~\ref{table:catalog} is published in its entirety in the electronic edition of the \textit{Astronomical Journal}.  A portion is shown here for guidance regarding its form and content.}
\end{deluxetable}

%TRUEMODE\else
%TRUEMODE\include{tab2_first20}
%TRUEMODE\fi

%TRUEMODE\ifincludetable
\begin{deluxetable}{lcccccccccccc}
\tabletypesize{\scriptsize}
\tablecolumns{13}
\rotate
\tablewidth{0pt}
\tablecaption{SDSS--Spectoscopic Catalog\label{table:sdsscatalog}}
\tablehead{ \colhead{} & 
\colhead{R.A.} & 
\colhead{Decl.} & 
\multicolumn{3}{c}{SDSS information} & 
\colhead{Targ.} & 
\colhead{} & 
\colhead{} & 
\colhead{} & 
\colhead{} & 
\colhead{} & 
\colhead{} \\
\colhead{ID} & 
\colhead{(J2000)} & 
\colhead{(J2000)} & 
\colhead{Field} & 
\colhead{Fiber} & 
\colhead{Rerun} & 
\colhead{Code} & 
\colhead{class} & 
\colhead{subclass} & 
\colhead{$z$} & 
\colhead{$\sigma(z)$} & 
\colhead{Warn.} & 
\colhead{$\chi^2/\nu$} \\
\colhead{(1)} & 
\colhead{(2)} & 
\colhead{(3)} & 
\colhead{(4)} & 
\colhead{(5)} & 
\colhead{(6)} & 
\colhead{(7)} & 
\colhead{(8)} & 
\colhead{(9)} & 
\colhead{(10)} & 
\colhead{(11)} & 
\colhead{(12)} & 
\colhead{(13)}}
\startdata
   10 &  17\phn17\phn50.76 & +60\phn17\phn01.4 & 354 & 262 & 51792 & \phn\phn\phn\phn\phn\phn\phn       64 &galaxy & \nodata & \phs0.07923 & 0.00002 &  0 & 1.31 \\
   67 &  17\phn18\phn15.67 & +60\phn08\phn22.0 & 353 & 612 & 51703 & \phn\phn\phn\phn\phn\phn\phn       64 &galaxy & \nodata & \phs0.15671 & 0.00001 &  0 & 1.38 \\
  100 &  17\phn19\phn44.06 & +60\phn02\phn44.6 & 354 & 298 & 51792 & \phn\phn\phn\phn\phn\phn\phn       64 &galaxy & \nodata & \phs0.09421 & 0.00001 &  0 & 1.20 \\
  107 &  17\phn18\phn53.16 & +60\phn05\phn41.0 & 354 & 299 & 51792 & \phn\phn\phn\phn\phn\phn\phn       64 &galaxy & \nodata & \phs0.13285 & 0.00001 &  0 & 1.30 \\
  155 &  17\phn19\phn44.88 & +59\phn57\phn06.9 & 354 & 290 & 51792 & \phn\phn\phn\phn\phn\phn\phn       64 &galaxy & \nodata & \phs0.06890 & 0.00001 &  0 & 1.34 \\
  159 &  17\phn20\phn07.90 & +60\phn01\phn59.7 & 354 & 285 & 51792 & \phn\phn\phn\phn\phn\phn\phn       64 &galaxy & \nodata & \phs0.07541 & 0.00002 &  0 & 1.22 \\
  255 &  17\phn20\phn02.11 & +59\phn42\phn40.8 & 353 & 628 & 51703 & \phn\phn\phn\phn\phn\phn\phn\phn        4 &galaxy & \nodata & \phs0.23937 & 0.00000 &  0 & 1.73 \\
  256 &  17\phn19\phn44.14 & +59\phn41\phn01.0 & 354 & 292 & 51792 & \phn\phn\phn\phn\phn\phn\phn       64 &galaxy & AGN & \phs0.12926 & 0.00001 &  0 & 2.03 \\
  258 &  17\phn20\phn09.50 & +59\phn40\phn30.1 & 354 & 289 & 51792 & \phn\phn\phn\phn\phn\phn\phn       64 &galaxy & \nodata & \phs0.02760 & 0.00001 &  0 & 1.55 \\
  324 &  17\phn20\phn47.14 & +59\phn33\phn14.5 & 366 & 338 & 52017 & \phn\phn\phn\phn\phn\phn\phn       96 &galaxy & \nodata & \phs0.07041 & 0.00002 &  0 & 1.29 \\
  325 &  17\phn20\phn41.40 & +59\phn32\phn44.3 & 353 & 636 & 51703 & \phn\phn  1048580 &QSO & broadline & \phs1.18430 & 0.00037 &  0 & 1.47 \\
  327 &  17\phn21\phn02.95 & +59\phn33\phn12.9 & 353 & 634 & 51703 & \phn\phn\phn\phn\phn\phn\phn       64 &galaxy & \nodata & \phs0.07014 & 0.00001 &  0 & 1.49 \\
  369 &  17\phn20\phn02.83 & +59\phn22\phn50.0 & 353 & \phn 39 & 51703 & \phn\phn\phn\phn\phn\phn\phn\phn        0 &star & F5 & $-$0.00053 & 0.00002 &  0 & 1.15 \\
  374 &  17\phn20\phn19.94 & +59\phn24\phn16.5 & 366 & 335 & 52017 & \phn\phn\phn\phn\phn\phn\phn       64 &galaxy & \nodata & \phs0.15423 & 0.00001 &  0 & 1.42 \\
  377 &  17\phn21\phn04.75 & +59\phn24\phn51.4 & 366 & 334 & 52017 & \phn 34603008 &QSO & broadline & \phs0.78576 & 0.00034 &  0 & 1.41 \\
  380 &  17\phn20\phn56.71 & +59\phn19\phn36.8 & 366 & 382 & 52017 & \phn\phn\phn\phn\phn\phn\phn       64 &galaxy & \nodata & \phs0.06710 & 0.00001 &  0 & 1.33 \\
  381 &  17\phn21\phn28.70 & +59\phn19\phn41.3 & 353 & \phn 40 & 51703 & \phn\phn\phn\phn\phn\phn\phn       64 &galaxy & AGN & \phs0.06544 & 0.00001 &  0 & 1.66 \\
  429 &  17\phn21\phn52.92 & +59\phn11\phn55.5 & 366 & 385 & 52017 & \phn\phn\phn\phn\phn\phn\phn       96 &galaxy & \nodata & \phs0.06593 & 0.00001 &  0 & 1.54 \\
  430 &  17\phn20\phn45.22 & +59\phn16\phn17.7 & 353 & \phn 35 & 51703 & \phn\phn\phn\phn\phn\phn\phn       64 &galaxy & \nodata & \phs0.15352 & 0.00001 &  0 & 1.54 \\
  436 &  17\phn21\phn56.93 & +59\phn13\phn57.2 & 366 & 390 & 52017 & \phn\phn\phn\phn\phn\phn\phn       64 &galaxy & \nodata & \phs0.06535 & 0.00001 &  0 & 1.26 \\
  437 &  17\phn20\phn55.01 & +59\phn11\phn16.8 & 353 & \phn 30 & 51703 & \phn\phn\phn\phn\phn\phn\phn       64 &galaxy & \nodata & \phs0.06581 & 0.00001 &  0 & 1.37 \\
\enddata
\tablecomments{(1) Row synchronous ID; (2) SDSS right ascension of spectroscopic target in units of hours, minutes, and seconds; (3) SDSS declination of spectroscopic target in units of degrees, arcminutes, and arcseconds; (4) SDSS field; (5) SDSS fiber number; (6) SDSS reduction run; (7) SDSS primTarget code (decimal); (8) spectroscopic classification; (9) spectroscopic subclassification; (10) redshift; (11) redshift error; (12) redshift warning, $=$0 is safe value; (13) $\chi^2$ per degree of freedom for fit. Table~\ref{table:sdsscatalog} is published in its entirety in the electronic edition of the \textit{Astronomical Journal}.  A portion is shown here for guidance regarding its form and content.  The electronic version of this catalog includes an entry for each row in table~\ref{table:catalog}, including null rows with no spectroscopic information. }
\end{deluxetable}

%TRUEMODE\else
%TRUEMODE\include{tab3_first20}
%TRUEMODE\fi

%TRUEMODE\ifincludetable
\begin{deluxetable}{lcccccccccccccc}
\tabletypesize{\scriptsize}
\tablecolumns{15}
\rotate
\tablewidth{0pt}
\tablecaption{Hectospec--Spectoscopic Catalog\label{table:mmtcatalog}}
\tablehead{ \colhead{} & 
\colhead{R.A.} & 
\colhead{Decl.} & 
\multicolumn{3}{c}{Hectospec information} & 
\colhead{Targ.} & 
\colhead{} & 
\colhead{} & 
\colhead{} & 
\colhead{} & 
\colhead{} & 
\colhead{} & 
\colhead{Vis.} & 
\colhead{Fix} \\
\colhead{ID} & 
\colhead{(J2000)} & 
\colhead{(J2000)} & 
\colhead{Field} & 
\colhead{Fiber} & 
\colhead{Rerun} & 
\colhead{Code} & 
\colhead{class} & 
\colhead{subclass} & 
\colhead{$z$} & 
\colhead{$\sigma(z)$} & 
\colhead{Warn.} & 
\colhead{$\chi^2/\nu$} &
\colhead{Flag} & 
\colhead{Flag} \\
\colhead{(1)} & 
\colhead{(2)} & 
\colhead{(3)} & 
\colhead{(4)} & 
\colhead{(5)} & 
\colhead{(6)} & 
\colhead{(7)} & 
\colhead{(8)} & 
\colhead{(9)} & 
\colhead{(10)} & 
\colhead{(11)} & 
\colhead{(12)} & 
\colhead{(13)} & 
\colhead{(14)} & 
\colhead{(15)}}
\startdata
   54 &  17\phn18\phn24.87 & +60\phn18\phn58.5 & 132 & 265 & 300 &  4 &galaxy & \nodata & \phs0.40845 & 0.00011 &  0 & 1.63 &  1 &  0 \\
   61 &  17\phn18\phn23.47 & +60\phn11\phn04.1 & 132 & 251 & 300 &  4 &QSO & broadline & \phs1.42971 & 0.00078 &  0 & 1.03 &  1 &  0 \\
   65 &  17\phn19\phn23.05 & +60\phn13\phn16.4 & 132 & \phn 36 & 300 &  2 &galaxy & \nodata & \phs0.31870 & 0.00003 &  0 & 1.12 &  1 &  0 \\
   66 &  17\phn19\phn26.87 & +60\phn13\phn03.7 & 132 & \phn 38 & 300 &  2 &galaxy & \nodata & \phs0.16146 & 0.00001 &  0 & 1.66 &  1 &  0 \\
   68 &  17\phn18\phn31.25 & +60\phn08\phn32.1 & 132 & 255 & 300 &  4 &star & manual & $-$0.00066 & 0.00011 &  0 & 0.91 &  1 &  1 \\
   72 &  17\phn19\phn01.99 & +60\phn12\phn18.4 & 132 & \phn\phn  4 & 300 &  4 &galaxy & \nodata & \phs0.17466 & 0.00004 &  0 & 1.53 &  1 &  0 \\
   74 &  17\phn18\phn41.04 & +60\phn15\phn07.2 & 132 & 263 & 300 &  4 &galaxy & manual & \phs0.21740 & 0.00011 &  0 & 1.31 &  1 &  1 \\
   76 &  17\phn18\phn42.88 & +60\phn10\phn00.9 & 132 & 266 & 300 &  4 &galaxy & \nodata & \phs0.27900 & 0.00004 &  0 & 1.23 &  1 &  0 \\
   79 &  17\phn18\phn45.97 & +60\phn17\phn06.5 & 132 & \phn\phn  2 & 300 &  2 &galaxy & \nodata & \phs0.21761 & 0.00003 &  0 & 1.04 &  1 &  0 \\
   81 &  17\phn19\phn27.99 & +60\phn17\phn17.6 & 132 & \phn 39 & 300 &  4 &galaxy & \nodata & \phs0.24725 & 0.00006 &  0 & 0.88 &  1 &  0 \\
   82 &  17\phn18\phn25.34 & +60\phn14\phn11.3 & 132 & 268 & 300 &  4 &galaxy & \nodata & \phs0.62651 & 0.00006 &  0 & 0.93 &  1 &  0 \\
   88 &  17\phn18\phn28.76 & +60\phn11\phn00.3 & 132 & 262 & 300 &  2 &galaxy & \nodata & \phs0.26943 & 0.00006 &  0 & 0.98 &  1 &  0 \\
   89 &  17\phn18\phn57.85 & +60\phn10\phn58.0 & 132 & 261 & 300 &  4 &galaxy & \nodata & \phs0.24753 & 0.00006 &  0 & 0.91 &  1 &  0 \\
   91 &  17\phn19\phn14.34 & +60\phn10\phn42.0 & 132 & \phn\phn  8 & 300 &  2 &galaxy & \nodata & \phs0.23165 & 0.00002 &  0 & 1.19 &  1 &  0 \\
   96 &  17\phn18\phn40.72 & +60\phn07\phn25.6 & 132 & 253 & 300 &  1 &star & F5 & $-$0.00110 & 0.00001 &  0 & 3.15 &  1 &  0 \\
   98 &  17\phn18\phn43.49 & +60\phn02\phn24.8 & 132 & 291 & 300 &  2 &galaxy & \nodata & \phs0.13305 & 0.00001 &  0 & 2.84 &  1 &  0 \\
   99 &  17\phn18\phn42.16 & +60\phn03\phn16.6 & 132 & 258 & 300 &  4 &galaxy & \nodata & \phs0.12986 & 0.00002 &  0 & 1.05 &  1 &  0 \\
  104 &  17\phn18\phn24.77 & +60\phn05\phn17.2 & 132 & 256 & 300 &  2 &galaxy & \nodata & \phs0.12923 & 0.00001 &  0 & 5.08 &  1 &  0 \\
  109 &  17\phn18\phn57.82 & +60\phn01\phn45.8 & 132 & 260 & 300 &  4 &galaxy & \nodata & \phs0.18926 & 0.00003 &  0 & 0.96 &  1 &  0 \\
  112 &  17\phn19\phn28.74 & +60\phn03\phn45.1 & 132 & \phn\phn  6 & 300 &  2 &galaxy & \nodata & \phs0.20937 & 0.00003 &  0 & 1.02 &  1 &  0 \\
  113 &  17\phn19\phn46.96 & +60\phn03\phn42.6 & 132 & \phn 31 & 300 &  2 &galaxy & \nodata & \phs0.27874 & 0.00001 &  0 & 1.09 &  1 &  0 \\
\enddata
\tablecomments{(1) Row synchronous ID; (2) Hectospec right ascension of spectroscopic target in units of hours, minutes, and seconds; (3) Hectospec declination of spectroscopic target in units of degrees, arcminutes, and arcseconds; (4) Hectospec field; (5) Hectospec fiber number; (6) Hectospec reduction run; (7) Hectospec Target code (decimal); (8) spectroscopic classification; (9) spectroscopic subclassification; (10) redshift; (11) redshift error; (12) redshift warning, $=$0 is safe value, $>$0 is safe only if Visual flag in (14) is =1; (13) $\chi^2$ per degree of freedom for fit.  (14) Visual inspection flag, 1 indicates secure redshift, 0 indicates dubious or insecure redshift; (15) Flag indicates automatic redshift discarded and replaced (see text).  Table~\ref{table:mmtcatalog} is published in its entirety in the electronic edition of the \textit{Astronomical Journal}.  A portion is shown here for guidance regarding its form and content. The electronic version of this catalog includes an entry for each row in table~\ref{table:catalog}, including null rows with no spectroscopic information.}
\end{deluxetable}

%TRUEMODE\else
%TRUEMODE\include{tab4_first20}
%TRUEMODE\fi

\end{document}